\documentclass[twocolumn,tighten]{aastex62}
\usepackage{natbib}
\usepackage{epsfig}
\usepackage{xcolor}

\newcommand{\acounits}{\mbox{M$_\odot$ pc$^{-2}$ (K km s$^{-1}$)$^{-1}$}}

\shorttitle{Forming Super Star Clusters in NGC 253}
\shortauthors{Leroy, Bolatto et al.}

\begin{document}

\title{Forming Super Star Clusters in the Central Starburst of NGC 253}

\author{Adam~K.~Leroy}
\affil{Department of Astronomy, The Ohio State University, 140 West 18th Avenue, Columbus, Ohio 43210, USA}

\author{Alberto D. Bolatto}
\affil{Department of Astronomy, University of Maryland, College Park, Maryland 20742, USA}

\author{Eve C. Ostriker}
\affil{Department of Astrophysical Sciences, Princeton University, Princeton, New Jersey 08544, USA}

\author{Fabian Walter}
\affil{Max-Planck-Institut f\"{u}r Astronomie, K\"{o}nigstuhl 17, D-69117, Heidelberg, Germany}

\author{Mark Gorski}
\affil{Department of Physics and Astronomy, University of Western Ontario, London, Ontario N6A 3K7, Canada}

\author{Adam Ginsburg}
\affil{National Radio Astronomy Observatory, PO Box O, 1003 Lopezville Road, Socorro, New Mexico 87801, USA}
\affil{Jansky Fellow}

\author{Nico Krieger}
\affil{Max-Planck-Institut f\"{u}r Astronomie, K\"{o}nigstuhl 17, D-69117, Heidelberg, Germany}

\author{Rebecca C. Levy}
\affil{Department of Astronomy, University of Maryland, College Park, Maryland 20742, USA}

\author{David S. Meier}
\affil{New Mexico Institute of Mining \& Technology, 801 Leroy Place, Socorro, New Mexico 87801, USA}
\affil{National Radio Astronomy Observatory, PO Box O, 1003 Lopezville Road, Socorro, New Mexico 87801, USA}

\author{Elisabeth Mills}
\affil{Department of Astronomy, Boston University, 725 Commonwealth Avenue, Boston, Massachusetts 02215, USA}

\author{J\"{u}rgen Ott}
\affil{National Radio Astronomy Observatory, PO Box O, 1003 Lopezville Road, Socorro, New Mexico 87801, USA}

\author{Erik~Rosolowsky}
\affil{Department of Physics, University of Alberta, Edmonton, AB T6G 2E1, Canada}

\author{Todd A. Thompson}
\affil{Department of Astronomy, The Ohio State University, 140 West 18th Avenue, Columbus, Ohio 43210, USA}
\affil{Center for Cosmology \& Astro-Particle Physics, The Ohio State University, Columbus, Ohio 43210}

\author{Sylvain Veilleux}
\affil{Department of Astronomy, University of Maryland, College Park, Maryland 20742, USA}

\author{Laura K. Zschaechner}
\affil{University of Helsinki, Physicum, Helsingin Yliopisto, Gustaf H\"{a}llstr\"{o}min katu 2, 00560 Helsinki, Finland}
\affil{Finnish Center for Astronomy with ESO}

\submitjournal{ApJ}

\begin{abstract}
NGC~253 hosts the nearest nuclear starburst. Previous observations show a region rich in molecular gas, with dense clouds associated with recent star formation. We used ALMA to image the $350$~GHz dust continuum and molecular line emission from this region at 2~pc resolution. Our observations reveal $\sim 14$ bright, compact ($\sim 2{-}3$ pc FWHM) knots of dust emission. Most of these sources are likely to be forming super star clusters (SSCs) based on their inferred dynamical and gas masses, association with $36$~GHz radio continuum emission, and coincidence with line emission tracing dense, excited gas. One source coincides with a known SSC, but the rest remain invisible in {\em Hubble} near-infrared (IR) imaging. Our observations imply that gas still constitutes a large fraction of the overall mass in these sources. Their high brightness temperature at $350$~GHz also implies a large optical depth near the peak of the IR spectral energy distribution. As a result, these sources may have large IR photospheres and the IR radiation force likely exceeds $L/c$. Still, their moderate observed velocity dispersions suggest that feedback from radiation, winds, and supernovae are not yet disrupting most sources. This mode of star formation appears to produce a large fraction of stars in the burst. We argue for a scenario in which this phase lasts $\sim 1$~Myr, after which the clusters shed their natal cocoons but continue to produce ionizing photons. The strong feedback that drives the observed cold gas and X-ray outflows likely occurs after the clusters emerge from this early phase.
\end{abstract}


\section{Introduction}
\label{sec:intro}

Vigorous bursts of star formation in galaxy centers and merging galaxies produce ``super'' star clusters \citep[SSCs, e.g.,][]{HOLTZMAN92,WHITMORE03,PORTEGIESZWART10}. The SSCs in well-known local starbursts like M82 and the Antennae galaxies have been studied for decades \citep[e.g.,][]{WHITMORE03,MCCRADY05}. These massive (M$_\star >10^5$\,M$_\odot$), compact ($R\sim1$\,pc) concentrations of stars may be younger cousins to the Milky Way's globular clusters.

Less extreme massive young stellar clusters ($M_\star \gtrsim 10^4$~M$_\odot$, age $\lesssim 100$~Myr) have been found in the Milky Way and many nearby galaxies and \citep[see review by][]{PORTEGIESZWART10,LONGMORE14}. Overall, the fraction of stars formed in clusters appears to increase with the surface density of star formation \citep[][]{KRUIJSSEN12,JOHNSON16,GINSBURG18B}. Because higher levels of star formation activity were prevalent at $z\sim1-3$, the formation of SSCs may represent a mode of star formation more common in the early universe than today. Studying the present-day formation of SSCs may thus offer a window into how star formation proceeded during the epoch of galaxy assembly \citep[e.g.,][]{ZHANG10}.

Gas structures associated with recent or future formation of SSCs have been identified in the Antennae galaxies \citep{HERRERA12,JOHNSON15} and the Large Magellanic Cloud \citep{OCHSENDORF17}. But despite decades of optical and near infrared studies, only a pair of candidate \textit{forming} SSCs have been resolved in cold gas and dust emission\citep{TURNER17,OEY17}. In both cases, CO~(3-2) emission has been seen associated with an SSC in a starburst dwarf galaxy. This CO~(3-2) emission may trace moderately more excited and dense gas than the CO~(1-0) line.

In the Milky Way, $\sim 4$ massive protocluster candidates have been identified \citep[e.g.,][]{GINSBURG12,FUKUI16,LONGMORE17,URQUHART18}, often following the criteria of \citet{BRESSERT12}. The Milky Way proto-clusters have gas mass $\lesssim 10^5$~M$_\odot$, somewhat lower than the extragalactic proto-SSC candidates. They appear likely to form $M_\star \sim 3 \times 10^4$~M$_\odot$ clusters, assuming $\sim 30\%$ efficiency \citep[see][]{BRESSERT12}.

In this paper, we report the identification of $14$ candidate proto-SSCs in NGC~253. This galaxy hosts one of the nearest nuclear starbursts \citep[$d \sim 3.5$~Mpc;][]{REKOLA05}, which produces stars at a rate of $\sim 2$~M$_\odot$~yr$^{-1}$ \citep[e.g.,][]{LEROY15A,BENDO15}. This burst is fed by the galaxy's strong bar \citep{SORAI00}, making NGC 253 a prototype for the common phenomenon of  bar-fed nuclear starbursts \citep[see][]{KORMENDY04}. 

\cite{WATSON96} and \citet{KORNEI09} used the \textit{Hubble} Space Telescope to discover a young ($\sim 6$~Myr) heavily extinguished ($A_V \sim 17$~mag) SSC in the nuclear region of NGC~253. Any other SSCs in the nuclear region must be too heavily embedded to appear prominent in {\em Hubble} images, including the near-infrared (IR) images presented in  \citet[][]{WALTER17}. But \citet{ULVESTAD97}, following \citet{TURNER85}, showed the presence of $\sim 30$ flat spectrum, compact ($\sim 1$~pc) radio sources that could be embedded {\sc Hii} regions. One of these coincides with the SSC of \citet{WATSON96} and \citet{KORNEI09}.

Previous mm- and submm-wave observations show that the NGC 253 nuclear region hosts massive, dense molecular clouds \citep{SAKAMOTO11,LEROY15A}. The whole region resembles a heavily scaled up version of the Milky Way's Central Molecular Zone \citep{SAKAMOTO11}. Observations at $\theta \approx 0.5''$ resolution show that these clouds host compact $< 10$~pc sized clumps of gas and dust, which have the appropriate masses to form massive clusters and are associated with signatures of massive star formation \citep{ANDO17}. Given these candidate SSC-forming structures and the presence of at least one known SSC, NGC 253 is the ideal target to catch SSC formation in the act.

In this paper, we use the Atacama Large submillimeter/Millimeter Array (ALMA) to map dust emission from the NGC 253 starburst at $0.11'' \approx 1.9$~pc resolution, a factor of $5$ improvement in linear scale (and a factor of 25 in area) compared to \citet{ANDO17}. This allows us resolve individual forming SSCs, which have sizes of a few pc \citep[e.g., see][]{PORTEGIESZWART10,BRESSERT12,LONGMORE14}, at the heart of the \citet{ANDO17} clumps. 

\section{Observations}
\label{sec:obs}

\begin{figure*}
\centering
\includegraphics[height=4.5in]{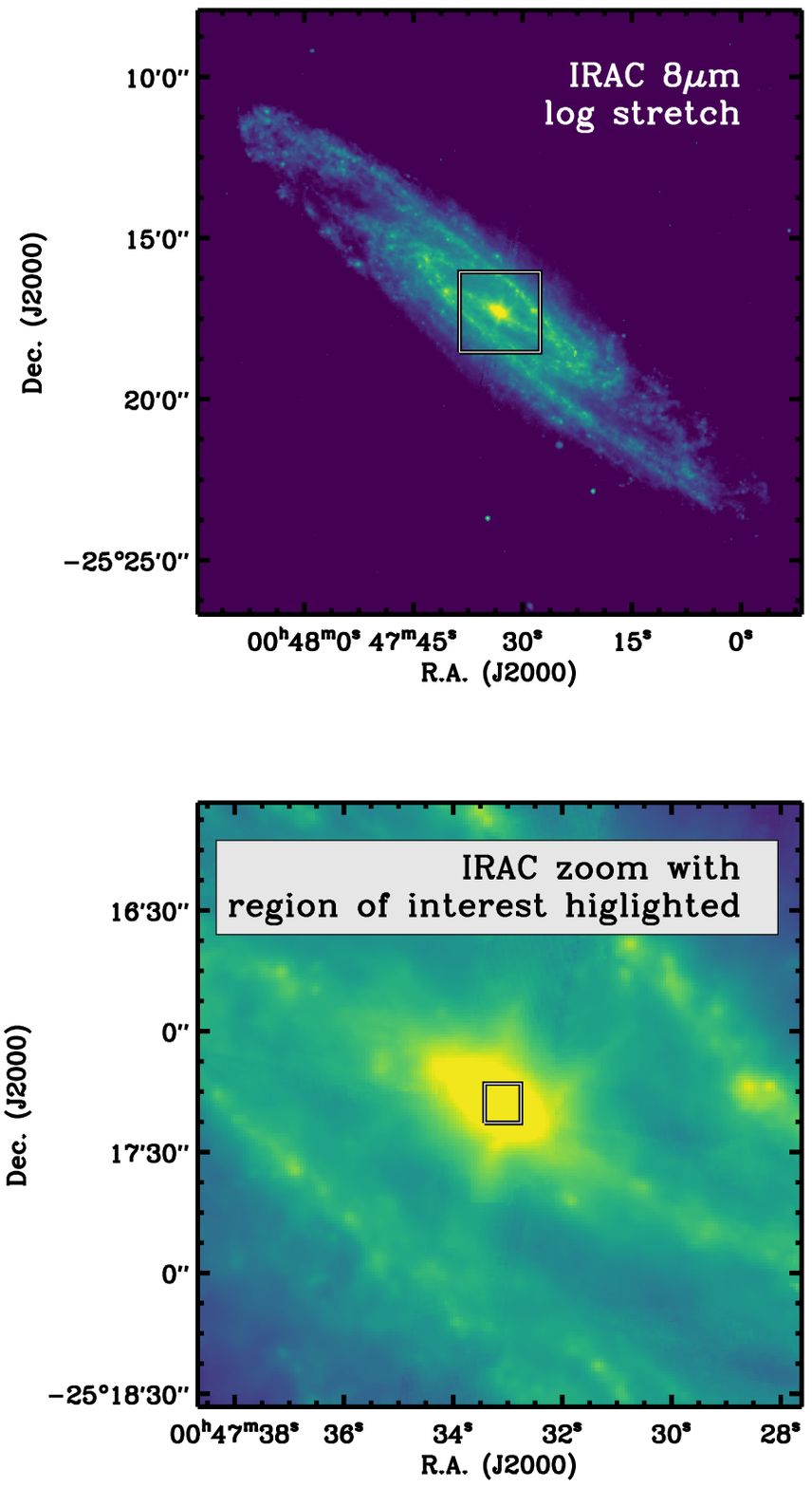}
\includegraphics[height=4.5in]{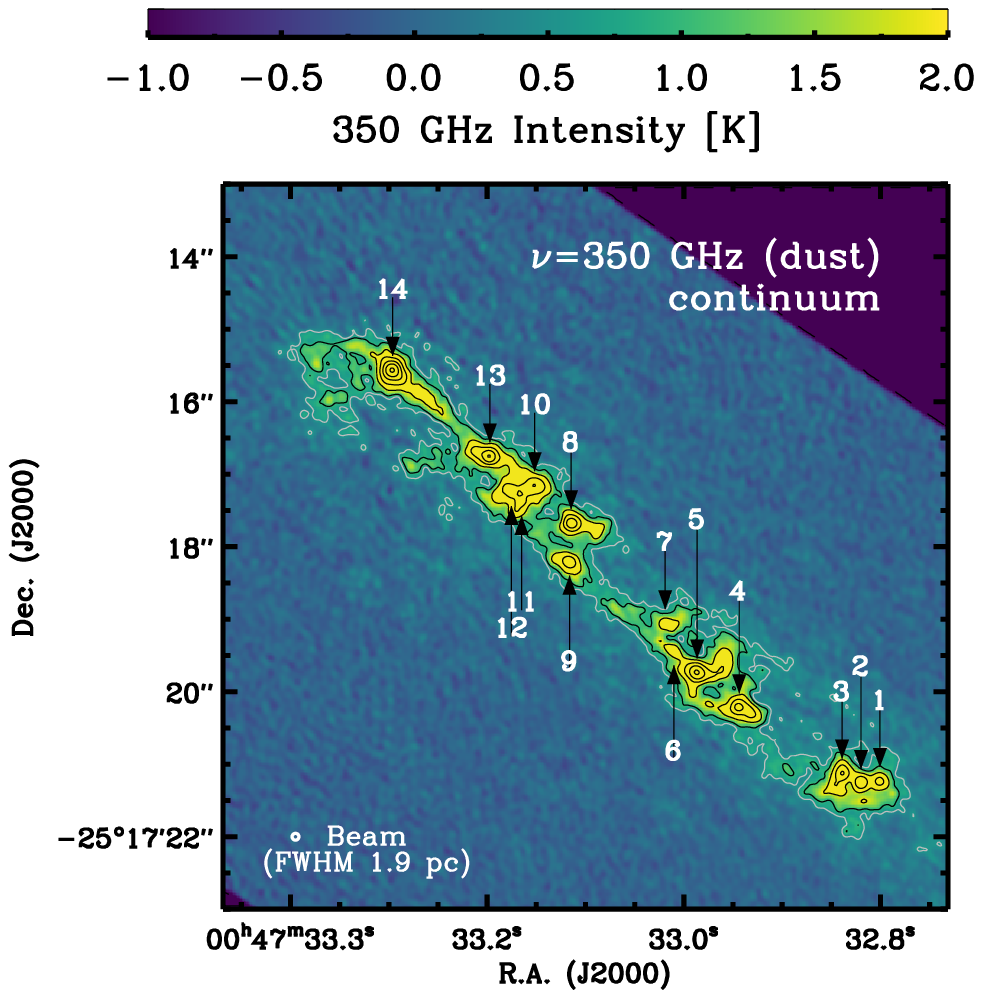}
\caption{({\em top left}) LVL IRAC 8$\mu$m image of NGC~253 \citep{DALE09,LEE09}. The square shows the region highlighted in the bottom left panel. ({\em bottom left}) The field analyzed in this paper (square box) plotted over the central part of the $8\mu$m image. We focus on the innermost region of the galaxy. This region hosts $\sim 10$ dense molecular clouds \citep[e.g.,][]{SAKAMOTO11,LEROY15A,ANDO17}, a large amount of high density gas \citep[e.g.,][]{PAGLIONE01,KNUDSEN07,LEROY15A,MEIER15}, and forms stars at a rate of $\approx 2~$M$_\odot$~yr$^{-1}$ \citep[see][]{BENDO15,LEROY15A}. ({\em right}) ALMA $\nu = 350$~GHz continuum emission from the inner region of NGC 253 at $0.11\arcsec \sim 1.9$~pc resolution. The emission, which is mostly from dust at this frequency, shows 14 bright peaks, each only moderately extended relative to the $1.9$~pc beam. The sizes, implied dust optical depths at $\nu = 350$~GHz ($\sim 850\mu$m), kinematics, and association with dense, excited gas suggest that many of these peaks may represent forming super star clusters (see Sections \ref{sec:candidates} and \ref{sec:props}). Note that we show a $10'' \times 10''$ cutout covering the region of interest, but that the FWHM of the primary beam of ALMA's 12\,m antennas is $18''$ at $350$~GHz. Contours in the continuum image show 0.6~K (gray), and 1, 2, 4, ... K (black).
\label{fig:continuum}}
\end{figure*}

\begin{figure*}
\centering
\plottwo{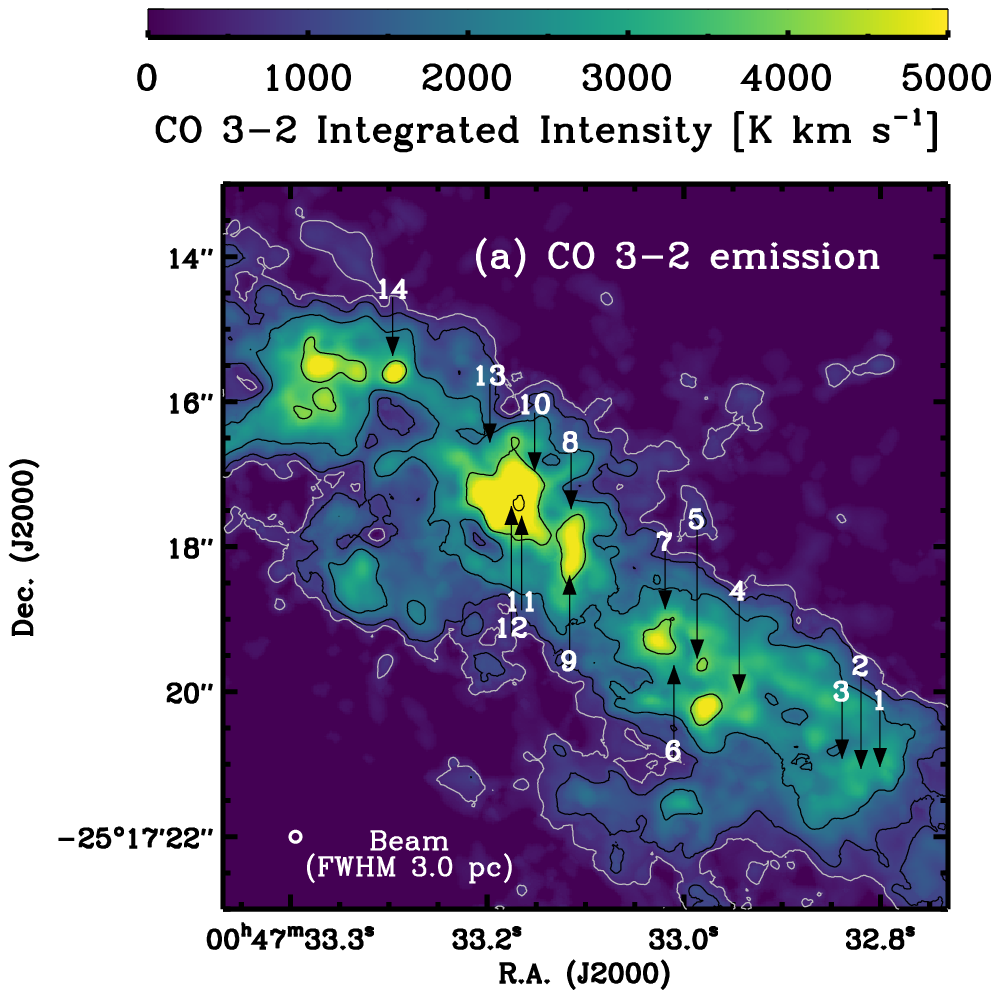}{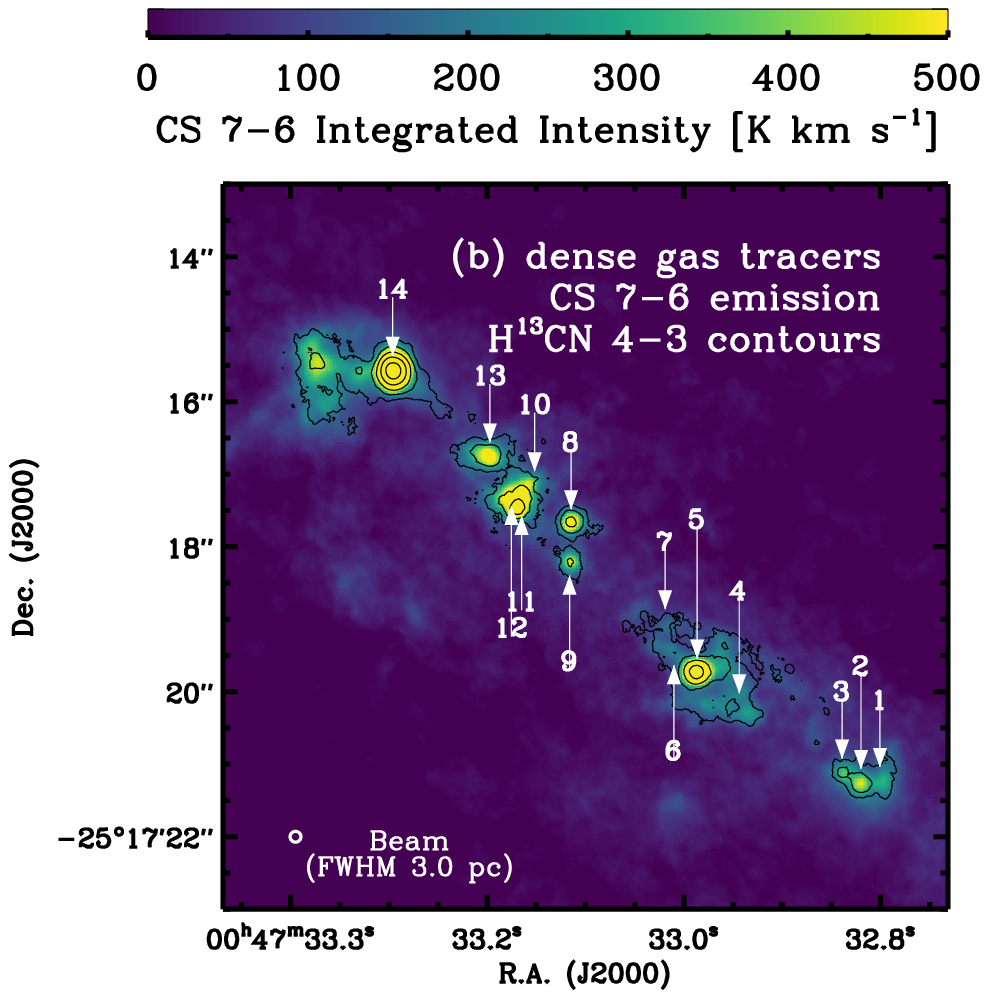}
\plottwo{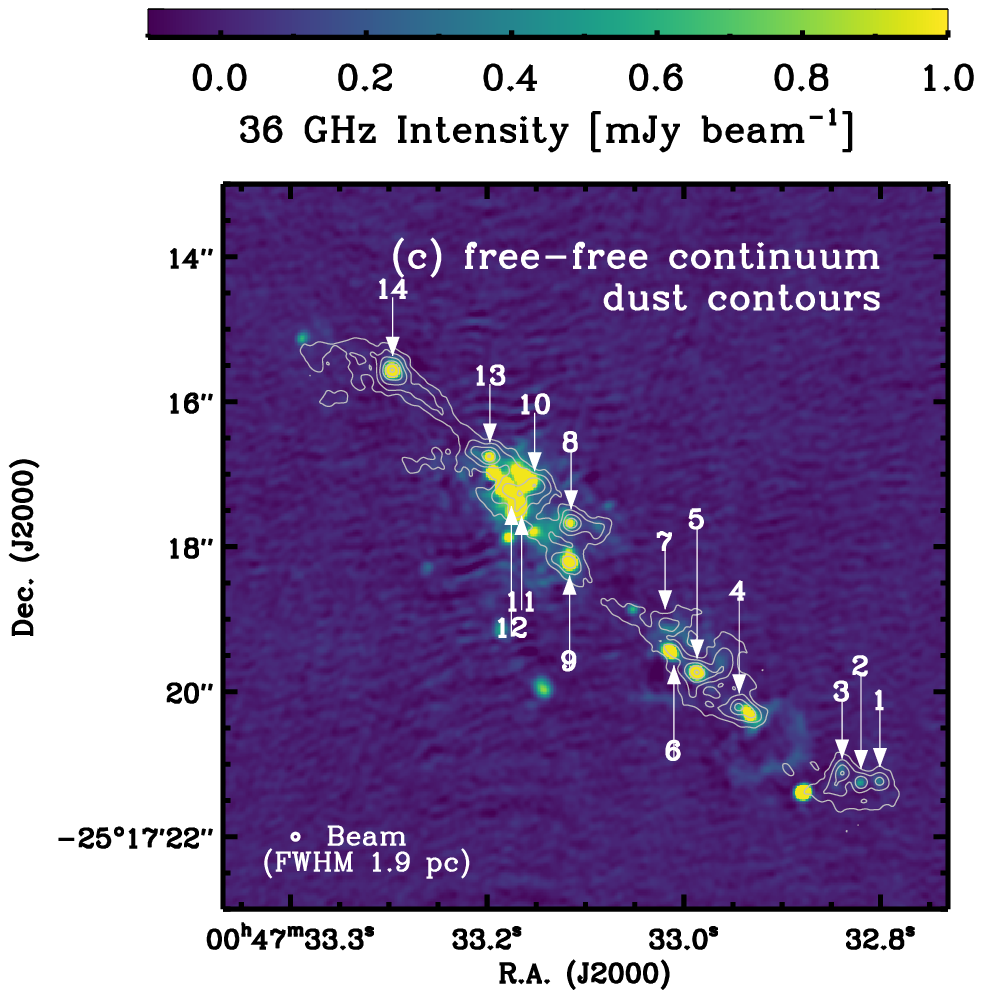}{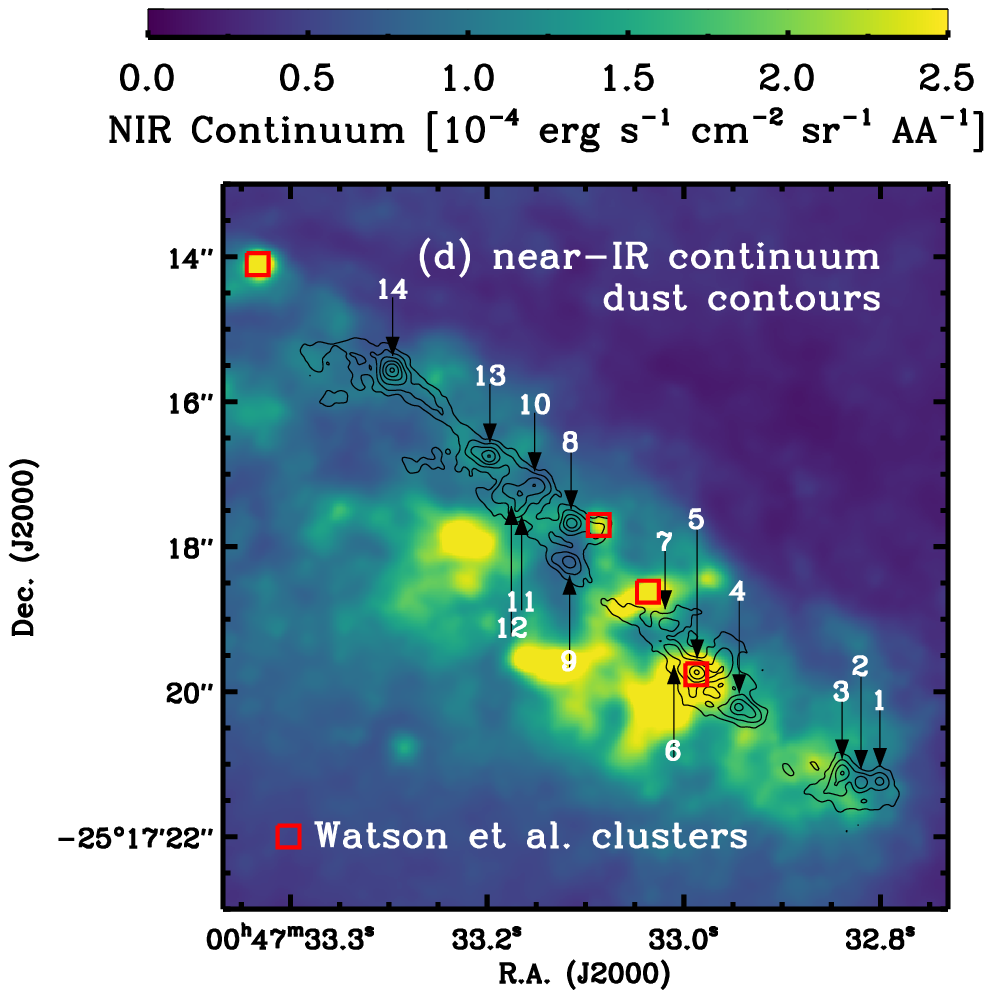}
\caption{Our region of interest in gas, radio continuum, and near-IR emission. The region is rich in molecular gas and signatures of recent star formation. Panel \textbf{(a)} shows a sprawling, high column density distribution of CO~(3-2) emission (N. Krieger et al. in preparation). However, the CO integrated intensity on its own only roughly suggests the dense peaks seen in the continuum images. On the other hand, panel \textbf{(b)} shows that the continuum sources do coincide with peaks of emission from CS~(7-6) and H$^{13}$CN~(4-3), tracers of dense, excited molecular gas. Panel \textbf{(c)} shows that the continuum peaks are also mostly coincident with bright $\nu = 36$~GHz radio continuum emission. Radio continuum at this frequency and resolution is most likely to be free-free, tracing ionizing photon production by young stars. Despite the prodigious concentrations of gas and the likely presence of embedded massive stars, these sources are mostly missing from the HST near-IR continuum image shown in panel \textbf{(d)}. In that panel, we mark the four clusters identified from HST imaging by \citet{WATSON96}. Among our candidate forming clusters, only source \# 5 appears prominent at $1.3\mu$m. It coincides with the previously known SSC from \citet{WATSON96} and \citet{KORNEI09}. Contour levels: (a) CO~(3-2) integrated intensity at 500~K~km~s$^{-1}$ (gray), then 1000, 2000, ... K~km~s$^{-1}$ (black); (b) H$^{13}$CN integrated intensity 50, 100, 200, ... (black); (c) ALMA 350~GHz continuum image (i.e., Figure \ref{fig:continuum}) at 1, 2, 4, ... K (gray); (c) ALMA 350~GHz continuum image (i.e., Figure \ref{fig:continuum}) at 1, 2, 4, ... K (black).
\label{fig:lines}}
\end{figure*}

We used ALMA to observe NGC 253 at $\nu \sim 350$~GHz ($\lambda \sim 850\mu$m) as part of project 2015.1.00274.S (P.I. A. Bolatto). We observed with the main array in both an intermediate configuration and a 2-km extended configuration. We also used 7-m array in the Atacama Compact Array (ACA) to recover short spacing information. The bandpass captures the sub-mm continuum from dust emission and covers several molecular rotational transitions, including CO~(3-2), HCN~(4-3), HCO$^+$~(4-3), CS~(7-6), and H$^{13}$CN~(4-3). The full suite of line images and the extended molecular gas distribution and kinematics traced by CO~(3-2) will be presented by N. Krieger et al. (in preparation).

We combined the observatory-provided calibrated visibilities for two 12-m configurations and the ALMA Compact Array (ACA) 7-m array and imaged them in version $5.1.1$ of the Common Astronomy Software Application (CASA) using CASA's \texttt{tclean} task. The inclusion of the ACA means that scales out to 19\arcsec\ are recovered.

We are interested in the compact structures at the heart of the starburst. Therefore, when imaging the continuum, we adopted a Briggs robust parameter $r=-2$ (i.e., nearly uniform weighting). This weights the extended baselines more heavily in order to produce a higher resolution image. For the lines of interest in this paper, CS~(7-6) and H$^{13}$CN~(4-3), sensitivity remains a concern. Therefore in the line images we emphasized surface brightness sensitivity and used a standard Briggs weighting with robust parameter $r=0.5$. 

After imaging, we convolved the continuum and line images to convert from an elliptical to a round beam shape. For the continuum image used in this paper, the fiducial frequency is $\nu = 350$~GHz and the final FWHM beam size is $\theta = 0.11\arcsec$. Before the convolution to a round beam, the beam of the continuum image has major and minor FWHM $0.105\arcsec \times 0.065\arcsec$.

The rms noise away from the source in the cleaned, round beam image is $0.2$~K in Rayleigh Jeans brightness temperature units, equivalent to $\approx 0.2$~mJy~beam$^{-1}$. For the H$^{13}$CN~(4-3) and CS~(7-6) line images the beam size is $0.175''$ (convolved from $\sim 0.14'' \times 0.11''$) and the typical rms in the cube is $0.4$~K per $5$~km~s$^{-1}$ channel. The ancillary CO~(3-2) and HCN~(4-3) observations have similar resolution and noise. More details of the line imaging appear in N. Krieger et al. (in preparation).

We compare the ALMA data to Karl G. Jansky VLA imaging of $\nu = 36$~GHz continuum emission \citep[][and M. Gorski et al. in preparation]{GORSKI17}. At this frequency and resolution, the radio continuum is likely to be predominantly free-free emission \citep[e.g., see][]{MURPHY11}. These data have native resolution slightly better than the ALMA continuum image, with a FWHM beam of $0.096\arcsec \times 0.45\arcsec$. We convolve them to the match the resolution of ALMA, $\theta = 0.11''$, for analysis. After convolution the VLA data have rms noise $\sim 0.03$~mJy~beam$^{-1}$.

We also compare to \textit{Hubble} Space Telescope imaging of the near-IR ($\lambda = 1.3\mu$m) continuum. These were obtained to serve as an off-line continuum measurement for the Paschen $\beta$ observations presented by \citet{WALTER17}.

\section{Candidate Forming Super Star Clusters}
\label{sec:candidates}

The top left panel in Figure \ref{fig:continuum} shows whole disk of NGC 253 seen at 8$\mu$m  by the Local Volume Legacy survey \citep[LVL][]{DALE09,LEE09}. The $8\mu$m image shows the location of UV-heated small dust grains (likely polycyclic aromatic hydrocarbons (PAHs)), and so illustrates the overall morphology of the ISM in the galaxy. The bottom left panel zooms in on the square region indicated in the top panel. The square in the bottom left panel shows our regions of interest in this paper. This is a square field, $10'' \times 10''$ across that includes most of the active star formation and dense clumps in the nuclear starburst. ALMA's 12\,m antennas have a primary beam of $18''$ at 350~GHz, so the ALMA observations cover a somewhat larger field of view than we show in the Figure. In total, the nuclear burst in NGC 253 contains $\sim 3 \times 10^8$~M$_\odot$ of molecular material and forms stars at $\sim 2$~M$_\odot$~yr$^{-1}$ \citep{LEROY15A,BENDO15}.

\subsection{Dust Continuum and Gas}
\label{sec:dustandgas}

The right panel of Figure \ref{fig:continuum} shows $\nu = 350$~GHz ($\lambda\sim 855~\mu {\rm m}$) continuum emission from this inner region at $\theta = 0.11\arcsec \approx 1.9$~pc resolution. At this frequency, thermal emission from large dust grains represents most of the emission, with $350$~GHz in the Rayleigh-Jeans part of the spectral energy distribution. In all but the most extreme conditions (which may include these peaks), this emission is optically thin. Thus, modulo temperature and emissivity variations, this emission offers an optically thin tracer of the column density distribution in the burst.

Our imaging reveals $14$ bright, compact continuum peaks embedded in a network of extended emission with brightness temperatures $T_b \sim 0.5{-}1$~K. We identify their locations using the local maximum finding routine from CPROPS \citep{ROSOLOWSKY06}. This program finds peaks that (1) exceed all other pixel values within a square search kernel 2 times the synthesized beam across, and are (2) at least $5\sigma$ above any contour shared with another peak, or $5\sigma$ above $0$~K if there is no shared contour.

These peaks have brightness of a few K up to a few tens of K and FWHM sizes of $\sim 2.5{-}4$~pc before any deconvolution. Thus, they appear bright and compact, but still marginally resolved by our beam. As we will see, these sizes and the implied gas and dust masses of $\sim 10^4{-}10^6$~M$_\odot$ suggest that these structures are forming SSCs.

The bright peaks are still associated with large surrounding reservoirs of gas. We show this in the top left panel of Figure \ref{fig:lines}, where we plot the line-integrated CO~(3-2) intensity. The peaks sit at the hearts of the massive clouds and clumps studied by \citet{SAKAMOTO11}, \citet{LEROY15A}, and \citet{ANDO17}. They are not conspicuous in the integrated CO~(3-2) intensity, although N. Krieger et al. (in preparation) show that they can be identified from the CO kinematically.

The continuum peaks stand out in lines that trace high density molecular gas. The top right panel of Figure \ref{fig:lines} shows our region of interest in line-integrated CS~(7-6) as a color image with contours showing line-integrated H$^{13}$CN (4-3) intensity. The H$^{13}$CN (4-3) line emits most effectively at densities $n\gtrsim 10^7$~cm$^{-3}$ \citep{SHIRLEY15} and $T\gtrsim40$~K. The CS~(7-6) emission, which also traces warm, dense gas, has critical density $\sim 3 \times 10^7$~cm$^{-3}$ and requires $T\gtrsim60$~K. 

At coarser resolution, HCN~(4-3) and CS~(7-6) emission correlate with IR emission in star-forming galaxies, with a linear relationship relating IR and line luminosity \citep[][]{ZHANG14,TAN18}. Here we see H$^{13}$CN~(4-3) and CS~(7-6) emission directly associated with the sites of massive star and cluster formation on $\sim 2$~pc scales. This direct association of these high density tracers with the sites of massive star formation fits in to a broader picture in which spectroscopic tracers of dense gas correlate with the rate of recent star formation \citep[e.g.,][]{GAO04}, with tracers of the densest gas showing the most linear correlations.

\subsection{Signatures of Massive Star Formation}
\label{sec:vla}

Are there actually signatures of young, massive stars associated with these peaks of gas and dust emission? The bottom left panel of Figure \ref{fig:lines} shows $\nu \sim 36$~GHz continuum emission from our target field \citep[][and M. Gorski et al. in preparation]{GORSKI17} with the dust continuum contours overlaid. Our dust continuum peaks are coincident with, or near to, peaks of bright radio continuum emission.

At this frequency and resolution, most of the sources in the $\nu = 36$~GHz map arise from free-free emission. Modulo loss of ionizing photons to dust and a mild dependence on the electron temperature and Gaunt factor, free-free emission directly traces ionizing photon production in a manner similar to optical recombination lines. But unlike optical and near-IR recombination line emission, $36$~GHz emission is almost totally unaffected by extinction. Thus the bottom left panel of Figure \ref{fig:lines} suggests that young, heavily embedded massive stars lie at or near most of our observed dust clumps.

The bottom right panel of Figure \ref{fig:lines} shows that these signatures of massive star formation are almost totally obscured by dust even in the near infrared. We plot the near infrared ($1.3\mu$m) continuum as seen by {\em Hubble}'s Wide Field Camera 3 (filter F130N). We also indicate the position of the four clusters identified by \citet{WATSON96} from earlier {\em Hubble} near-IR imaging. To match the astrometry of our near-IR data, which aligns well with the ALMA and VLA observations, we found it necessary to shift the measured positions from \citet{WATSON96} by $\Delta \alpha , \Delta \delta \approx +0.32\arcsec, -0.5\arcsec$.

The image shows bright stellar continuum emission coincident with the brightest SSC known from \citet{WATSON96} and \citet{KORNEI09}. Otherwise our dust peaks do not correspond to clear enhancements in the near-IR continuum. Given the presence of free-free emission, these sources are likely to be bright, compact, massive collections of young stars. But the extinction in the inner region of the galaxy is too severe to pick them out even in the near-infrared. 
This overwhelming extinction is striking, but not surprising. From the top left panel in Figure \ref{fig:lines}, we see that our peaks all lie at $I_{CO 3-2} \gtrsim 2,000$~K~km~s$^{-1}$. Under the conservative assumptions of thermalized CO lines, a low $\alpha_{\rm CO} = 0.8$~\acounits , and a Galactic dust-to-gas ratio, this amount of gas still corresponds to $E(B-V) \sim 34$~mag, or $A_J \sim 30$~mag. Even without accounting for the dense concentrations within the clouds, the central region of NGC 253 is heavily extinguished and capable of hiding luminous clusters at near-IR wavelengths.

This 36~GHz view of the NGC~253 nucleus resembles the $\sim 43$~GHz, 3~pc resolution view of M82 by \citet{TSAI09}. In M82, another starburst at $d \sim 3.5$~Mpc, \citealt{TSAI09} observed $\sim 9$ compact continuum sources likely to be heavily embedded {\sc Hii} regions powered by massive clusters. \citealt{TSAI09} showed these compact {\sc Hii} regions to exist within dense gas structures observed at $\sim 45$~pc resolution. Based on our observations of NGC~253, it seems plausible, even likely, that some of the individual \citet{TSAI09} sources will still be in the process of formation and that high resolution sub-mm observations of M82 would show associated pc-scale concentrations of gas and dust. \citet{SCHINNERER07} observed similar coincidence of dense gas tracers and continuum signatures of embedded massive star formation at $\sim 10$~pc resolution in the inner region of NGC~6946, though there were some detailed differences between their HCN map and the NGC~6946 continuum emission seen by \citet{TSAI06}. \citet{TURNER03} and \citet{TURNER04} found similar compact {\sc Hii} regions surrounding the SSC in NGC~5253. \citet{TURNER17} showed CO emission from the same source, though that emission appears optically thin in CO, perhaps indicating that the NGC 5253 cluster is at a later evolutionary stage than the ones that we observe.

\begin{figure*}
\centering
\plottwo{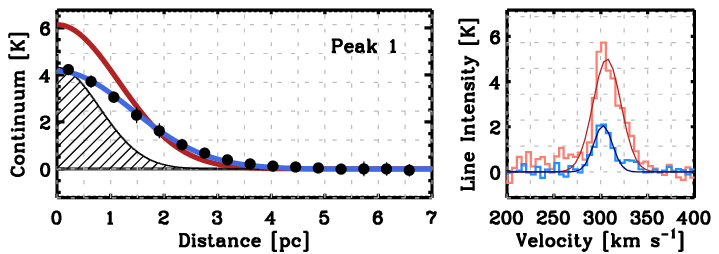}{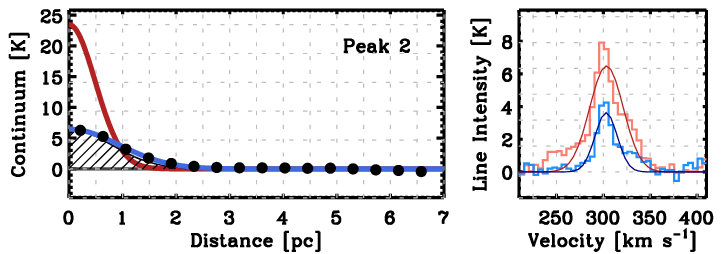}
\plottwo{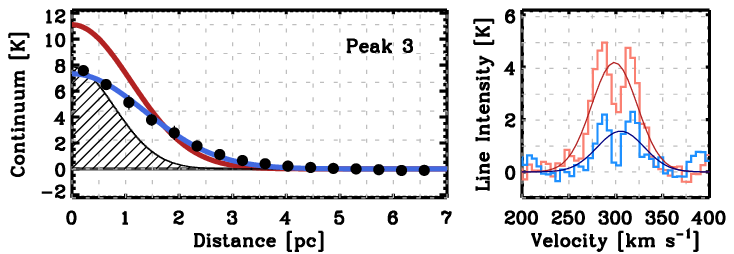}{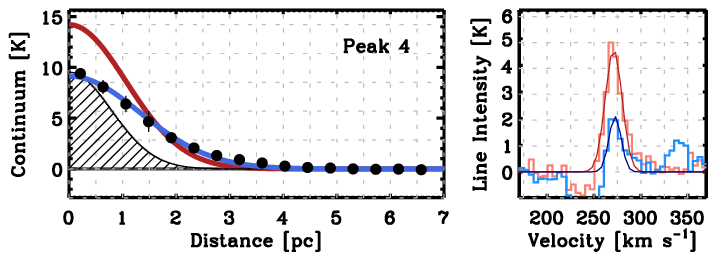}
\plottwo{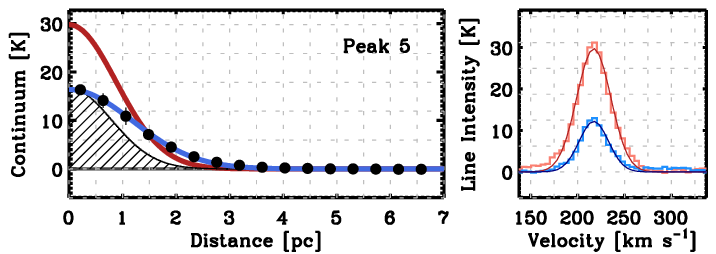}{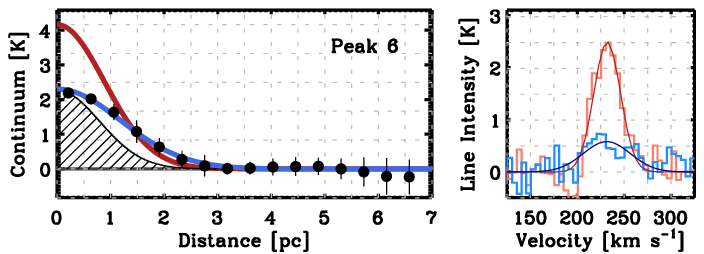}
\plottwo{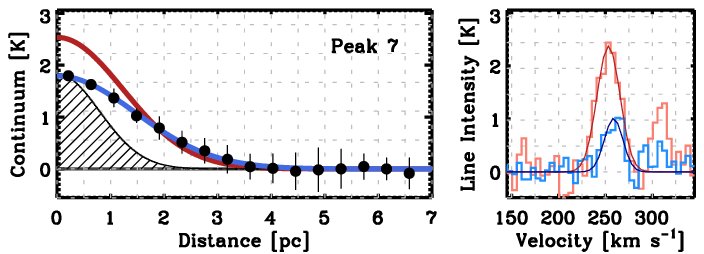}{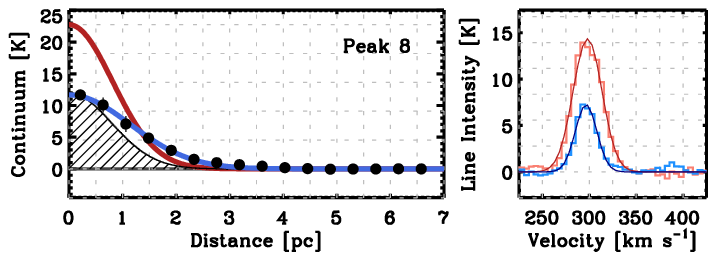}
\plottwo{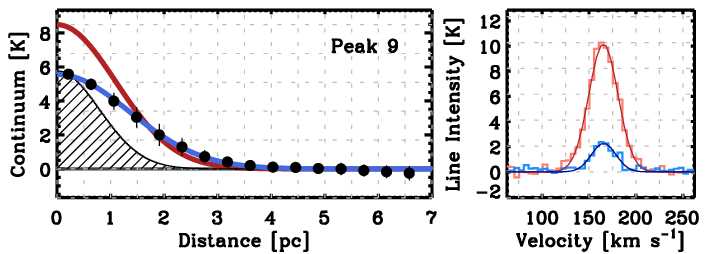}{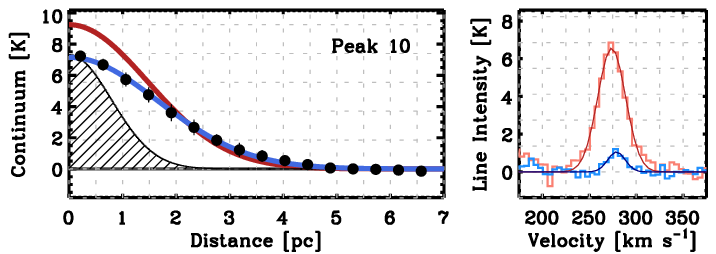}
\plottwo{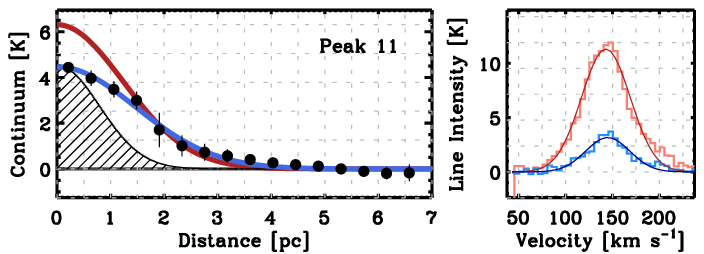}{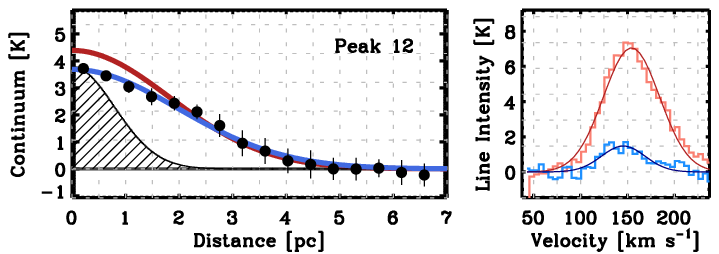}
\plottwo{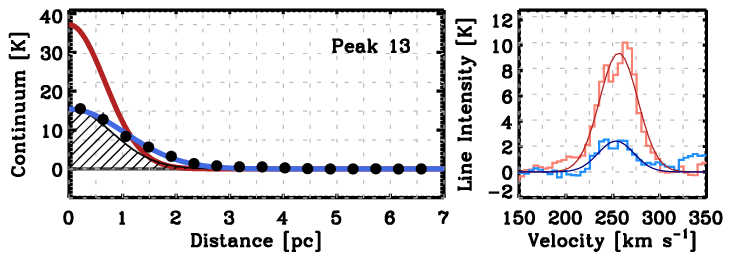}{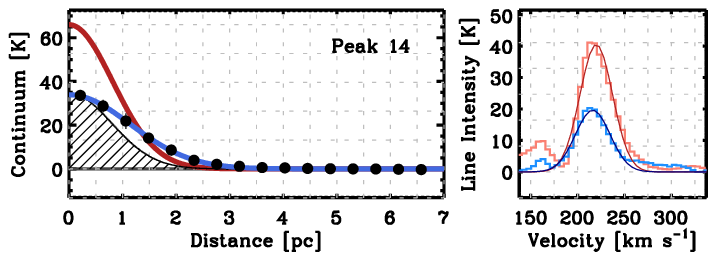}
\caption{Spatial ({\em left}) and spectral ({\em right}) profiles of our $14$ peaks. The left column shows the binned median-based radial profile of $350$~GHz emission about each peak (black bins, with a blue Gaussian fit). The black, shaded profile in each panel indicates the shape of the synthesized beam. The red profile shows the inferred shape of the peak after deconvolving the beam. The spectra show the background subtracted CS (7-6) emission in red and H$^{13}$CN (4-3) emission in blue. Lines indicate Gaussian fits to the profiles. Spectrum \#3  shows a split line profile, indicative of a shell geometry, self-absorption, or substructure. \label{fig:profiles}}
\end{figure*}

\begin{figure}
\centering
\plotone{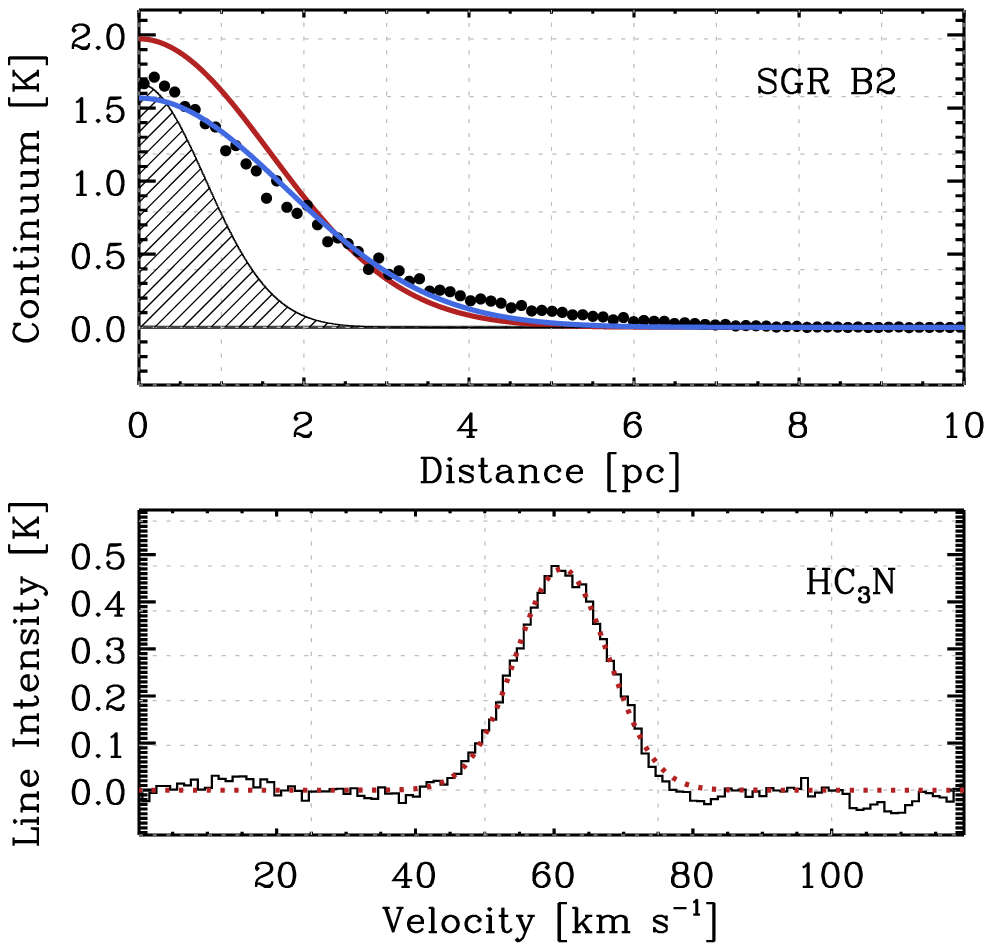}
\caption{Profile similar to those in Figure \ref{fig:profiles} but for the known Milky Way protoclusters in Sgr B2 \citep[e.g., see][]{GINSBURG18}. We show the profile of the sources as seen by ATLASGAL \citep[see][]{URQUHART18} matched to our $1.9$~pc resolution (top). On the bottom we show a spectrum of HC$_3$N~(24-23). In both cases, we have plotted the data similar to how we show our profiles, removing a local background from the radial profile and subtracting the continuum from the spectrum. The combined Sgr B2 protoclusters show peak continuum brightness $\sim 2$~K, FWHM extent $\sim 3.7$~pc, and $1\sigma$ line width $\sim 6.7$~km~s$^{-1}$. Thus, they would appear as among the least bright and least compact of our sources, with the narrowest line widths. But they do overall resemble the sources that we see in NGC~253, so that our observations  pick out sources that resemble scaled up versions of known protoclusters in the Milky Way. Similar to Sgr B2, we might expect some of our sources to break up into two or more protoclusters within our $1.9$~pc beam.
\label{fig:sgrb2}}
\end{figure}

\section{Properties of the Candidate Forming Super Star Clusters}
\label{sec:props}

We estimate the size, line width, and fluxes associated with each peak and use these to gauge the masses of the candidate proto-SSCs in several ways.

\subsection{Size, Line Width, and Flux Measurements}
\label{sec:fluxes}

\begin{deluxetable*}{lccccccccc}
\tabletypesize{\scriptsize}
\tablecaption{Measured Properties Candidate Young Clusters in NGC 253 \label{tab:peaks}}
\tablewidth{0pt}
\tablehead{
\colhead{\#} & 
\colhead{R.A.} & 
\colhead{Dec.} & 
\colhead{$T_{\rm pk}$} &
\colhead{FWHM} & 
\colhead{$\sigma_{\rm v}$} &
\colhead{$F_{350}$} &
\colhead{$F_{\rm 350,app}$\tablenotemark{a}}&
\colhead{$F_{36,app}$\tablenotemark{a}} &
\colhead{$f_{\rm ff}$\tablenotemark{a}} \\ 
\colhead{} & 
\colhead{($^\circ$)} & 
\colhead{($^\circ$)} &
\colhead{(K)} &
\colhead{(pc)} &
\colhead{(km~s$^{-1}$)} & 
\colhead{(mJy)} &
\colhead{(mJy)} & 
\colhead{(mJy)} & 
\colhead{} 
}
\startdata
1 & 11.886669 & -25.289232 &  6.1$\pm  1.4$ & 2.7$\pm 0.3$ & 12.4$\pm  2.4$ &  11.0$\pm  1.1$ & 14.52$\pm  1.45$ &  0.51$\pm  0.05$ & 0.03$\pm 0.01$ \\
2 & 11.886749 & -25.289240 & 23.5$\pm 12.2$ & 1.2$\pm 0.3$ & 14.9$\pm  3.0$ &   7.5$\pm  0.8$ & 12.06$\pm  1.21$ &  0.55$\pm  0.06$ & 0.04$\pm 0.01$ \\
3 & 11.886829 & -25.289200 & 11.1$\pm  2.6$ & 2.6$\pm 0.3$ & 24.8$\pm  0.3$ &  18.7$\pm  1.9$ & 24.40$\pm  2.44$ &  0.32$\pm  0.03$ & 0.01$\pm 0.01$ \\
4 & 11.887265 & -25.288950 & 14.2$\pm  3.4$ & 2.5$\pm 0.3$ &  7.7$\pm  1.2$ &  21.0$\pm  2.1$ & 31.48$\pm  3.15$ &  2.57$\pm  0.26$ & 0.07$\pm 0.02$ \\
5 & 11.887444 & -25.288816 & 29.8$\pm  8.7$ & 2.1$\pm 0.3$ & 16.9$\pm  1.2$ &  27.8$\pm  2.8$ & 42.81$\pm  4.28$ &  6.86$\pm  0.69$ & 0.13$\pm 0.02$ \\
6 & 11.887542 & -25.288733 &  4.2$\pm  1.3$ & 2.1$\pm 0.3$ & 19.7$\pm  5.4$ &   1.7$\pm  0.2$ &  6.24$\pm  0.62$ &  4.91$\pm  0.49$ & 0.63$\pm 0.40$ \\
7 & 11.887579 & -25.288628 &  2.5$\pm  0.6$ & 2.9$\pm 0.3$ & 11.0$\pm  1.2$ &   5.0$\pm  0.5$ & 10.64$\pm  1.06$ &  0.87$\pm  0.09$ & 0.07$\pm 0.05$ \\
8 & 11.887978 & -25.288244 & 22.7$\pm  7.1$ & 1.9$\pm 0.3$ & 13.4$\pm  2.0$ &  20.0$\pm  2.0$ & 28.47$\pm  2.85$ &  1.65$\pm  0.16$ & 0.05$\pm 0.02$ \\
9 & 11.887984 & -25.288389 &  8.5$\pm  2.0$ & 2.6$\pm 0.3$ & 14.3$\pm  1.2$ &   9.6$\pm  1.0$ & 19.82$\pm  1.98$ &  8.02$\pm  0.80$ & 0.32$\pm 0.10$ \\
10 & 11.888132 & -25.288099 &  9.2$\pm  1.6$ & 3.5$\pm 0.3$ & 12.1$\pm  2.8$ &  24.7$\pm  2.5$ & 36.57$\pm  3.66$ &  4.85$\pm  0.48$ & 0.11$\pm 0.06$ \\
11 & 11.888187 & -25.288160 &  6.3$\pm  1.3$ & 2.9$\pm 0.3$ & 24.7$\pm  1.3$ &   7.9$\pm  0.8$ & 18.86$\pm  1.89$ &  9.81$\pm  0.98$ & 0.41$\pm 0.11$ \\
12 & 11.888230 & -25.288122 &  4.4$\pm  0.7$ & 4.3$\pm 0.3$ & 26.6$\pm  3.3$ &   5.9$\pm  0.6$ & 29.99$\pm  3.00$ & 26.72$\pm  2.67$ & 0.71$\pm 0.23$ \\
13 & 11.888322 & -25.287989 & 37.1$\pm 14.2$ & 1.6$\pm 0.3$ & 19.7$\pm  0.7$ &  22.2$\pm  2.2$ & 32.82$\pm  3.28$ &  1.66$\pm  0.17$ & 0.04$\pm 0.02$ \\
14 & 11.888734 & -25.287657 & 66.0$\pm 20.6$ & 1.9$\pm 0.3$ & 18.0$\pm  0.5$ &  56.8$\pm  5.7$ & 85.05$\pm  8.50$ &  7.43$\pm  0.74$ & 0.07$\pm 0.01$ \\
\enddata
\tablenotetext{a}{Values measured in apertures centered on the peaks. The apertures have radius equal to the FWHM size of the source before deconvolution (i.e., to recover this add $1.9$~pc in quadrature to the value in the table). See text for more details.}
\tablecomments{R.A., Dec. refer to peak position in the $350$~GHz continuum image. FWHM size assumes a distance of $3.5$~Mpc and already accounts for the deconvolution of the $0.11\arcsec \approx 1.9$~pc beam. $T_{\rm pk}$ reports the peak intensity, after deconvolving the beam, at $350$~GHz in Rayleigh Jeans brightness temperature units. $\sigma_v$ is the linear average of the CS~(7-6) and H$^{13}$CN~(4-3) rms line width. $F_{350}$ is the flux at $350$~GHz estimated from the Gaussian fit to the profile. $F_{36}$ is the flux of $36$~GHz emission obtained from aperture photometry, this is scaled into a luminosity using the distance and used to estimate $Q_0$ and $M_\star$ using the equations in the text. $f_{\rm ff}$ refers to the estimated fractional free-free contribution to the $350$~GHz emission based on comparing fluxes measured in matched apertures.}
\end{deluxetable*}

\textbf{Size, Peak Temperature, and Flux at $350$~GHz:} To measure the sizes associated with each peak, we build an azimuthally averaged radial profile centered on the peak. In each half-beam thick ring centered on the peak, we calculate the median intensity. Using the median suppresses the influence of nearby peaks and the bright surrounding filamentary features, and so emphasizes the profile of the central peak. We further reject $3\sigma$ outliers about this median profile. Figure \ref{fig:profiles} shows the resulting profiles appear as black points, with error bars showing the scatter about the profile. Blue lines show a Gaussian fit to these profiles; this fit includes a background term, which is small in all cases. 

To compute deconvolved sizes we subtract the beam size in quadrature from the Gaussian fit to the profile. We also correct the peak temperature for the effects of the beam by scaling the measured peak temperature by the ratio of measured source area to the deconvolved source area. The deconvolved profiles appear as red lines in Figure \ref{fig:profiles}. 

We report the measured sizes in Table \ref{tab:peaks}. As a check, we also fit two dimensional Gaussians to each source. The deconvolved FWHM size from the Gaussian fits agree with our measured sizes with a scatter of $\pm 0.3$~pc. We adopt this as our uncertainty on the size, with the uncertainty dominated by the choice of methodology. We take the fractional uncertainty in the deconvolved $T_{\rm peak}$ to be the sum in quadrature of the fractional uncertainty due to statistical noise and the fractional uncertainty in the deconvolved beam area.

From the fits to the profiles, we also calculate the flux of each source at $\nu=350$~GHz, which we give in Table \ref{tab:peaks}. The statistical uncertainties on this flux are low, because the peaks are all detected at high signal to noise. In this case, we quote a $10\%$ uncertainty on the overall flux for each source, reflecting a mixture of calibration uncertainty (which should be covariant among all sources), uncertainty in the image reconstruction, and uncertainty due to the adopted methodology.

For reference, from a lower resolution, robust-weighted version of the continuum map, we calculate a total 350~GHz flux of $\approx 1.9$~Jy by integrating all emission above $S/N = 3$. The sources in Table \ref{tab:peaks} have total flux $0.28$~Jy, and so account for $\sim 15\%$ of the total $350$~GHz emission from the nuclear region.

\textbf{Line Widths:} We measure the line width associated with each peak. To do this, we define a series of apertures. The aperture associated with a peak has radius equal to the FWHM fit (not deconvolved) size of the peak and sits centered on the peak. The background region associated with each aperture extends from radius $1$ to $3$ times the FWHM fit size of the central source. The background excludes apertures associated with other peaks. We calculate the source spectrum by subtracting the average spectrum in the background from the average spectrum in the aperture. Note that the central apertures that we use are never less than $0.22\arcsec$ across (diameter). Thus the measurement region is always at least moderately extended compared to the $0.175''$ beam of the H$^{13}$CN~(4-3) and CS~(7-6) cubes.

We fit a Gaussian to the background-subtracted for CS~(7-6) and H$^{13}$CN~(4-3) spectra, fitting over a velocity range picked by eye to cover the emission line. Figure \ref{fig:profiles} shows both background-subtracted spectra and the fits for each peak. We take the linear average of the two line widths as the characteristic line width for the source. We adopt one half the difference in the line width derived between the two lines as our uncertainty.

\textbf{36~GHz fluxes:} We measure fluxes for each source from the VLA 36~GHz map. To do this, we subtract the average intensity in the local background region from the region inside the aperture. Then we sum the flux inside the aperture. We use the same apertures used to derive the line widths. 

Similar to the case of the ALMA fluxes, the statistical uncertainty in the 36~GHz flux is small (compare the fluxes in Table \ref{tab:peaks} to the 0.02~mJy~beam$^{-1}$ noise). We adopt an uncertainty of $10\%$ to reflect calibration and image reconstruction uncertainties.

Based on the 36~GHz emission, we estimate the fractional contribution of free-free flux to the ALMA band via:

\begin{equation}
f_{\rm ff} \approx \left( \frac{36}{350} \right)^{0.1} \frac{F_{36}}{F_{350}}
\end{equation}

\noindent where the first factor reflects the expected $-0.1$ spectral index from optically thin free-free emission and $F_{36}$ and $F_{350}$ refer to the observed total flux at $36$~GHz and $350$~GHz. For this application only, we measure fluxes from the ALMA 350~GHz map in exactly the same way that we measure the 36~GHz fluxes (i.e., using aperture photometry and the same aperture definitions). We report both sets of fluxes in Table \ref{tab:peaks}.

$f_{\rm ff}$ is the fraction of the $350$~GHz flux in the aperture that can be attributed to free-free emission, assuming that all of the $36$~GHz flux comes from optically thin free-free emission. A value $\ll 1$ is expected if thermal dust emission makes a large contribution to the emission from the $350$~GHz band. A value $\sim 1$ indicates either that free-free emission contributes a large fraction of the 350~GHz emission or that the $36$~GHz emission is not free-free in nature (expected, e.g., if synchrotron contributes heavily).

We see high $f_{\rm ff}$ around four sources: peaks \# 6 ($f_{\rm ff} = 0.63$), 9 ($0.32$), 11 ($0.41$), and 12 ($0.71$). With only two bands, we cannot distinguish between contamination of the $350$~GHz band by free-free or the $36$~GHz band by synchrotron. Additional high resolution measurements at $\sim 100{-}200$~GHz and at $\sim 1{-}25$~GHz will help resolve the nature of the emission \citep[some observations at slightly coarser resolution already exist][]{MOHAN05,ULVESTAD97}. 

Peak \#6 is indeed weak in the ALMA map but a clear point source in the 36~GHz map. This may represent a cluster in a later stage of evolution or a supernova remnant. The other three sources lie near the galaxy nucleus. Peaks \#11 and 12 lie in a region where \citet{MOHAN05} do see substantial radio recombination line flux, but also complex structure. Those authors speculate that the nucleus, which is $\sim 10$~pc away might contribute to ionization in the region. In any case, we apply $f_{\rm ff}$ as a correction to the gas mass estimates, viewing this as the most conservative option.

\subsection{Resemblance to a Known Milky Way Protocluster} 

As a check, we construct profiles similar to those in Figure \ref{fig:profiles} for the known Galactic protoclusters Sgr B2. This pair of bright sources near the Galactic center is regarded as very likely to be forming young massive clusters \citep[see][and references therein]{GINSBURG18,URQUHART18}. We degrade the ATLASGAL 500$\mu$m data to a resolution of $1.9$~pc (FWHM) and scale the intensity assuming a spectral index of $4$ (i.e., optically thin dust with $\beta=2$). We also extract a spectrum of HC$_3$N~(24-23) at 3~pc (FWHM) resolution to serve as a proxy for our H$^{13}$CN and CS measurements, though note that HC$_3$N~(24-23) has larger excitation requirements than the H$^{13}$CN~(4-3) or CS~(7-6). 

The resulting profile and spectrum, shown in Figure \ref{fig:sgrb2}, show that Sgr~B2 would have a slightly larger size and narrower line width than our candidate clusters. It would also have among the lowest brightness temperatures of our sources. But overall, the structure in Figure \ref{fig:sgrb2} does resemble what we see for our NGC~253 sources (Figure \ref{fig:profiles}). The comparison gives us confidence that we detect moderately more compact, scaled up versions of a known Galactic protocluster.

Sgr~B2 appears as a single extended source in this exercise, but also breaks into two massive protoclusters at higher resolution \citep[e.g., see Figure 1 in][]{GINSBURG18}. Therefore, this comparison also highlights the likelihood that despite our high (for extragalactic work) $1.9$~pc resolution, some of our sources may break into two or more smaller, more compact protoclusters when observed at higher resolution. From a first look at $\sim 2$ times higher resolution ALMA observations obtained during review of this paper, many of our sources do resolve into several smaller structures at higher resolution. In almost all cases, though, a single bright source still contributes most of the sub-mm flux. While we might expect the sizes of our sources to shrink some and to find some nearby lower mass clusters, we expect our main results to hold with improved resolution.

\subsection{Gas, Stellar, and Dynamical Masses}
\label{sec:masses}

\begin{deluxetable*}{lcccccccc}
\tabletypesize{\scriptsize}
\tablecaption{Estimated Physical Properties Candidate Young Clusters in NGC 253 \label{tab:masses}}
\tablewidth{0pt}
\tablehead{
\colhead{\#} & 
\colhead{$\log_{10} M_{\rm VT}$\tablenotemark{a}} &
\colhead{$\log_{10} M_{\rm gas}$\tablenotemark{b}} &
\colhead{$\log_{10} M_{\star}$\tablenotemark{c}} &
\colhead{$\log_{10} \Sigma_{\rm gas+\star}$\tablenotemark{b,c}} &
\colhead{$\log_{10} \rho_{\rm gas+\star}$\tablenotemark{b,c}} &
\colhead{$\log_{10} \tau_{\rm ff}$\tablenotemark{b,c}} &
\colhead{$p_r/M_\star$} &
\colhead{$v_{\rm esc}$} \\
\colhead{} & 
\colhead{(M$_\odot$)} &
\colhead{(M$_\odot$)} & 
\colhead{(M$_\odot$)} & 
\colhead{(M$_\odot$~pc$^{-2}$)} & 
\colhead{(M$_\odot$~pc$^{-3}$)} &
\colhead{(yr)} &
\colhead{(km~s$^{-1}$)} &
\colhead{(km~s$^{-1}$)}
}
\startdata
1 &  5.6 &  4.9 &  4.3 &  3.9 &  3.4 &  5.2 &  79.4 & 13.2 \\
2 &  5.4 &  4.7 &  4.3 &  4.6 &  4.4 &  4.7 &  64.7 & 18.2 \\
3 &  6.2 &  5.1 &  4.1 &  4.1 &  3.7 &  5.1 & 443.8 & 16.7 \\
4 &  5.1 &  5.1 &  5.0 &  4.4 &  3.9 &  4.9 &  18.6 & 22.0 \\
5 &  5.7 &  5.3 &  5.4 &  4.8 &  4.5 &  4.7 &  20.1 & 33.2 \\
6 &  5.9 &  3.6 &  5.3 &  4.5 &  4.1 &  4.9 &   0.8 & 21.8 \\
7 &  5.5 &  4.5 &  4.5 &  3.7 &  3.2 &  5.3 &  17.9 & 10.7 \\
8 &  5.5 &  5.2 &  4.8 &  4.6 &  4.2 &  4.8 &  51.3 & 23.3 \\
9 &  5.7 &  4.7 &  5.5 &  4.5 &  4.1 &  4.9 &   3.5 & 26.5 \\
10 &  5.7 &  5.2 &  5.3 &  4.3 &  3.7 &  5.1 &  17.1 & 22.4 \\
11 &  6.2 &  4.5 &  5.6 &  4.5 &  4.0 &  4.9 &   3.5 & 26.8 \\
12 &  6.4 &  4.1 &  6.0 &  4.6 &  3.9 &  5.0 &   0.5 & 35.3 \\
13 &  5.7 &  5.2 &  4.8 &  4.8 &  4.5 &  4.6 &  89.4 & 27.4 \\
14 &  5.7 &  5.7 &  5.5 &  5.1 &  4.8 &  4.5 &  53.3 & 45.5 \\
\enddata
\tablenotetext{a}{Dynamical mass from the virial theorem. Dominant uncertainty is sub-beam structure, including whether the source breaks up into multiple smaller sources, $\sim 0.3$~dex systematic uncertainty is plausible with our estimates likely to be high. Statistical uncertainty $\approx 0.1$~dex.}
\tablenotetext{b}{Gas-mass based quantities. Uncertainties from the assumed dust temperature, dust-to-gas ratio, and opacity. Likely magnitude is $\sim 0.4{-}0.5$~dex with our estimates likely to be low.}
\tablenotetext{c}{Zero age main sequence stellar mass needed to produce the observed 36~GHz emission as free-free following Equations \ref{eq:q0} and \ref{eq:mstar}. Mild uncertainty due to the assumed temperature, Gaunt factor, and possible contamination by synchrotron. Larger uncertainties regarding the amount of ionizing photons absorbed by dust and the possibility of pre-main sequence stars in the source.}
\tablecomments{Masses estimated following Section \ref{sec:masses}. The uncertainty in all of the mass estimates are dominated by systematic uncertainties. We note the dominant uncertainty for each quantity in the associated footnote. $M_{\rm VT}$ refers to the dynamical mass estimated from the virial theorem. $M_{\rm gas}$ refers to gas mass estimated from dust emission at $350$~GHz. $M_\star$ refers to the zero age main sequence stellar mass needed to match the ionizing photon production rate estimated from the 36~GHz emission. $\Sigma_{\rm gas+\star}$ is the estimated gas plus stellar surface density within the 2-d FWHM given our size and mass estimates. $\rho_{\rm gas+\star}$ is mass volume density within the 3-d FWHM given our given our size and mass estimates. $\tau_{\rm ff}$ is the gravitational free-fall time implied by that density. $p_r/M_\star$ is the equivalent radial momentum per unit stellar mass calculated from the gas velocity dispersion, gas mass, and stellar mass (Equation \ref{eq:radmom}).}
\end{deluxetable*}

\begin{figure*}
\centering
\plottwo{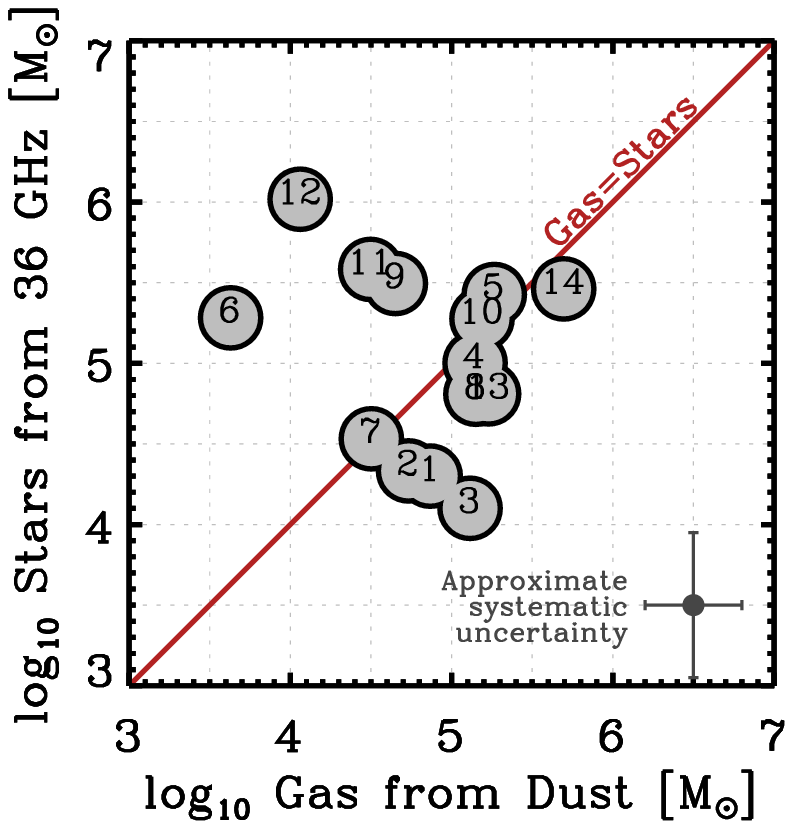}{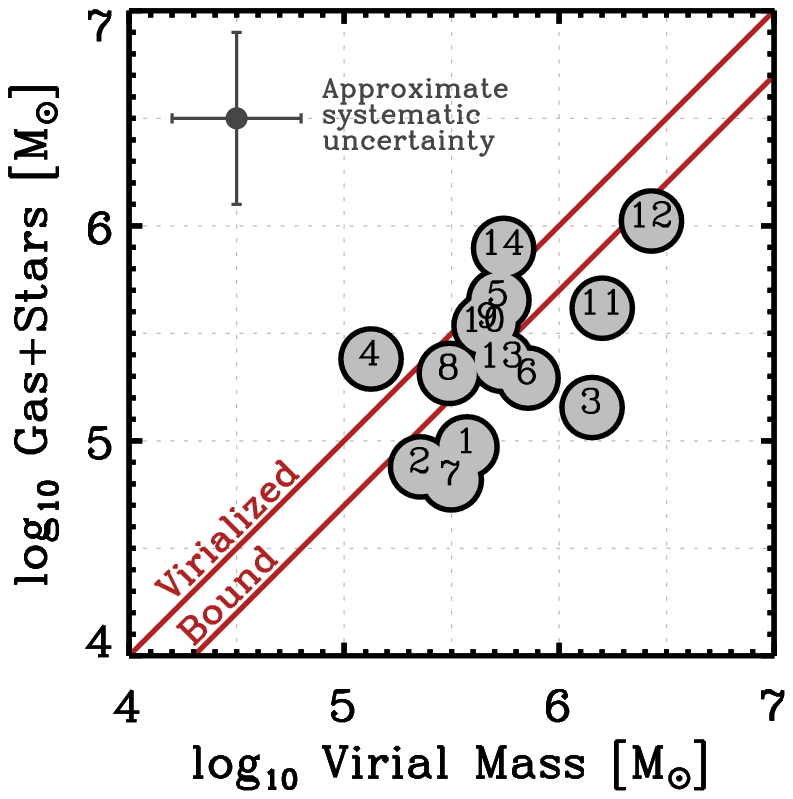}
\caption{Estimated mass budget in our candidate forming clusters. ({\em left}) Stellar mass ($y$-axis), estimated from the 36~GHz continuum emission assuming that it all arises from free-free emission and is produced by a population of stars on the zero age main sequence, as a function of gas mass ($x$-axis), estimated from the ALMA-observed dust continuum. ({\em right)} Combined gas plus stellar mass ($y$-axis) as a function of dynamical mass ($x$-axis) estimated from the measured size and line width of each source. In both panels the bold line shows equality with dotted lines offset by successive factors of two. The sources show a range of apparent gas richness, but most show at least half of their mass in gas. Given the signatures of dense, compact gas, these structures seem likely to still be forming. On average, we estimate a dynamical mass a factor of $\sim 2$ higher than the stellar and gas mass. This suggests either non-equilibrium contributions to the line width, a non-virialized dynamical state, or that some of our assumptions underestimate the true mass of gas and stars present.\label{fig:masses}}
\end{figure*}

Based on their size, line width, and fluxes, we estimate the gas, stellar, and dynamical masses of these cluster candidates. We report these in Table \ref{tab:masses}. To do this, we assume that (1) the free-free corrected 350~GHz emission arises from dust with some adopted temperature and emissivity, which we take to be well mixed with the gas with some characteristic dust-to-gas ratio; (2) the 36~GHz represents free-free emission emitted by a young stellar population on the zero age main sequence (ZAMS); and (3) our objects are in virial equilibrium, so that their sizes and line widths together indicate their dynamical mass.

\subsubsection{Zero Age Main Sequence Stellar Mass} 

Assuming that all of the $36$~GHz emission is produced by free-free interactions, we can estimate the ionizing photon production rate of each source. From this, we can calculate the mass of ZAMS stars needed to produce this number of ionizing photons.

Following \citet{MURPHY11} and \citet{CAPLAN86}, a $36$~GHz luminosity, $L_{\rm 36}$, implies an ionizing photon production rate of:

\begin{equation}
\label{eq:q0}
Q_0 \sim 1.06 \times 10^{26}~L_{\rm 36}~{\rm s}^{-1}~.
\end{equation}

\noindent Where we have assumed an electron temperature $T_e = 7,000$~K \citep[slightly higher than the estimate for NGC 253 by][]{BENDO15}. Here $Q_0$ is the ionizing photon production rate per second and $L_{\rm 36}$ is measured in erg~s$^{-1}$~Hz$^{-1}$.

Based on Starburst99 calculations \citep{LEITHERER99}, for a ZAMS population the mass ($M_{\star}$) relates to the ionizing photon production rate ($Q_0$) via

\begin{equation}
\label{eq:mstar}
M_{\star} \sim \frac{Q_0}{4 \times 10^{46}}~M_\odot~.
\end{equation}

\noindent We arrive at this value by simulating $10^6$~M$_\odot$ single stellar population, with the initial mass function of \citet{KROUPA01}, a maximum stellar mass of 100~$M_\odot$, and the default stellar evolution tracks and tuning parameters. Then we divide the ionizing photon output at time zero by mass of the stellar population. Together, Equations \ref{eq:mstar} and \ref{eq:q0} yield the mass of the embedded stellar population needed to produce the observed 36~GHz flux via free-free emission.

Our candidate proto-SSCs have median $\log_{10} M_\star [M_\odot] = 5.1$ and range  $\log_{10} M_\star [M_\odot] \sim 4.1{-}6.0$. Based on this calculation, all of our sources already qualify as young massive clusters \citep[i.e., $M_\star \gtrsim 10^4$~M$_\odot$][]{PORTEGIESZWART10,LONGMORE14}.

Note that the two highest values, for peaks \# 11 and 12, should be regarded with suspicion because of the high $f_{\rm ff}$ found for these objects (Table \ref{tab:peaks}) and the uncertain nature of the ionization in this region (see above). 

Our $M_\star$ is a linear translation of the 36~GHz flux, and our sources are detected at high signal-to-noise. This yields small statistical uncertainties, $\sim \pm 10\%$, and systematic effects dominate the uncertainty in $M_\star$. First, uncertainties in the Gaunt factor imply a systematic uncertainty of $\approx \pm20\%$ \citep{RYBICKI86}. Second, if dust absorbs a significant fraction of the ionizing photons, our $M_\star$ will represent an underestimate. Loss of ionizing photons to dust already appears to be a significant effect in massive star forming regions in the Milky Way \citep{BINDER18}, so we do expect this to also be important in the dustier, dense nucleus of NGC 253. But the magnitude of the effect is not clear. More, if there is ongoing accretion and many of the stars in the cluster have not yet reached the main sequence, we would also expect a higher mass per ionizing photon produced. Most of these effects have the sense that our quoted $M_\star$ likely represents a moderate underestimate.  The calculation also has the usual uncertainties related to the upper mass cutoff and shape of the IMF. Our estimates also take no account of the influence of binary stars \citep[e.g.,][]{ELDRIDGE17}. Note that, following \citet{XIAO18}, we expect the inclusion of binaries to have the largest effect after a few Myr. Thus we expect binarity to be a minor concern for this paper, which focuses on young sources. Finally, note that we estimate ZAMS mass in an aperture centered on the cluster. Future high resolution comparison of the $36$~GHz structure, dust, and gas will help us understand how much of this mass is, in fact, directly associated with the gas and dust peaks. 

\subsubsection{Implications for the Ionized Gas Content}

From $Q_0$ and a plausible size, we estimate the ionized gas mass and density associated with each source. We posit an {\sc Hii} region with radius $r_S$ at the heart of each source. Assuming Case B recombination, complete ionization of hydrogen, and $1.36$ contribution of helium by mass, we expect

\begin{eqnarray}
M_{\rm ion} &\approx& \left( \frac{Q_0}{\alpha_{\rm B}}~4/3 ~\pi~r_S^3 \right)^{0.5}~1.36 m_{\rm H} \\
\nonumber M_{\rm ion} &\approx& 684 M_\odot \left( \frac{Q_0}{10^{51}~{\rm s}^{-1}} \right)^{0.5}~\left( \frac{r_S}{{\rm 1~pc}} \right)^{1.5}~.
\end{eqnarray}

\noindent Here $\alpha_B \approx 3.4 \times 10^{-13}$~cm$^{3}$~s$^{-1}$ is the adopted recombination rate coefficient corresponding to an {\sc Hii} region temperature of 7000~K \citep{DRAINE11}. A higher ionizing photon flux requires more ionized gas to be present, and a larger {\sc Hii} region implies more ionized gas mass. The order of magnitude for our $M_{\rm ion}$ agrees with the calculations by \citet{ULVESTAD97}, though the adopted distances and other details do vary.

Measuring the sizes of the {\sc Hii} regions will help constrain this measurement, and is a natural next direction. For $r_S \sim 1$~pc, ionized gas contributes appreciably only to source \# 6 ($M_{\rm ion} / M_{\rm gas} \sim 50\%$ for our fiducial assumptions), source \# 11 ($\sim 10\%$), and source \# 12 ($\sim 40\%$). In all other cases, the fractional contribution of ionized gas to the gas mass is $< 10\%$ and usually $\lesssim 1\%$. Again, this implies that sources \# 11 and 12, which lie near the nucleus, need more detailed study \citep[see also][]{MOHAN05}.

The densities implied by this calculation appear reasonable.

\begin{equation}
n_{\rm ion} \approx 4850~{\rm cm}^{-3}~\left( \frac{Q_0}{10^{51}~{\rm s}^{-1}} \right)^{0.5}~
\left( \frac{r_S}{{\rm 1~pc}} \right)^{-1.5}
\end{equation}

\noindent yields densities mostly in the range $n_{\rm ion} \sim10^3-10^4$~cm$^{-3}$ for $r_S = 1$~pc. But this depends strongly on $r_S$. A measured value of $r_S$ will allow us to determin if the {\sc Hii} regions are overpressured and evolving \citep[and, e.g., to compare to][]{KRUMHOLZ09C} or in approximate pressure equilibrium with the surrounding gas \citep[e.g., as in the center of IC 342][]{MEIER11}.

\subsubsection{Gas Mass From Dust}

We estimate the mass of gas associated with each protocluster candidate from the 350~GHz dust emission. To do this, we estimate the optical depth at the peak by contrasting the measured brightness with an estimate of the true dust temperature. Then we convert the optical depth to a dust column using an assumed mass absorption coefficient. We convert the dust to a gas column via an adopted dust-to-gas ratio. Finally, we scale the gas column at the center of the proto-SSC by the area of the peak to calculate a total gas mass. In the future, we hope to constrain the dust-to-gas ratio by comparing our dust mass estimates to gas mass estimates based on the gas emission lines. At the moment, we consider the dust a more reliable estimate of the gas content than the molecular lines that we observe.

We assume a fiducial temperature of $T_{\rm dust} = 130$~K, assuming that the clusters coincide with the warm component seen in ammonia spectroscopy \citep{GORSKI17,MANGUM13} and that the gas and dust are collisionally coupled. We convert our measured 350~GHz intensity at the peak, $I_{350}$, to a dust optical depth via

\begin{equation}
\label{eq:tau}
I_{350} = \left[ 1 - \exp \left( -\tau_{\rm 350 GHz} \right) \right] B_\nu \left( T_{\rm dust} \right)~.
\end{equation}

\noindent Here $I_{350}$ is our measured 350~GHz intensity, corrected for free-free contamination using the value in Table \ref{tab:peaks}, and expressed in cgs units. $B_\nu$ is the Planck function evaluated at 350~GHz for our adopted dust temperature. This formulation deals better with mild optical depth effects than assuming the emission to be optically thin. However, if the emission is strongly clumped within our beam then these optical depth effects will be underestimated.

Equation \ref{eq:tau} yields optical depths at 350~GHz mostly in the range $0.035$ to $0.35$, with $\tau_{\rm 350 GHz} \sim 0.09$ on average. The dust appears to be moderately optically thin at 350~GHz. Because of the frequency-dependent opacity of dust, $\tau_{\nu} \propto \nu^\beta$ with $\beta \sim 1.5{-}2$, these values imply that these sources will be quite optically thick at higher frequencies (shorter wavelengths) where most of the energy is emitted.

After calculating $\tau$, we convert to a dust mass surface density using an assumed mass absorption coefficient, $\kappa$. We adopt  $\kappa = 1.9$~cm$^2$~g$^{-1}$. According to \citet{OSSENKOPF94}, this should be appropriate for $\nu \sim 350$~GHz and dust mixed with gas at density $\sim 10^5-10^6 ~{\rm cm^{-3}}$, but this value is uncertain by a factor of $\sim 2$.

Finally, we combine the dust surface density with an adopted dust-to-gas mass ratio, DGR, of 1-to-100, approximately the Milky Way value and similar to the value found for starburst galaxies by \citet{WILSON08B}. Then our estimate of the central gas surface density for each peak is:

\begin{equation}
\Sigma_{\rm gas} = \frac{1}{{\rm DGR} \kappa_{\rm 350 GHz}}~\tau_{\rm 350 GHz}
\end{equation}

We then scale this $\Sigma_{\rm gas}$ by the physical area of the peak, assuming each source to be a two dimensional Gaussian with the size quoted in Table \ref{tab:peaks}. Thus, $M_{\rm gas} = A \Sigma_{\rm gas}$.

We report the the results in Table \ref{tab:masses}. We find median $\log_{10} M_{\rm gas} [\rm{M}_\odot ] \sim 5.0$ and values in the range $\log_{10} M_{\rm gas} [\rm{M}_\odot ] \sim 3.6{-}5.7$.

As with $M_\star$, $M_{\rm gas}$ represents a nearly linear transformation of the measured source flux at $350$~GHz. Because we detect the sources at high S/N, the statistical uncertainties are quite low. Systematic uncertainties in the adopted temperature, mass absorption coefficient, and dust to gas ratio dominate the error budget for $M_{\rm gas}$.

Based on the excitation requirements of the lines that we see, and on the global spectral energy distribution (SED), $T_{\rm dust}$ seems unlikely to be lower than $\sim 50{-}60$~K in these dense, heated regions. Because the clusters are likely to be optically thick near the peak of the IR SEDs we can ask what temperature, along with our measured sizes, would place all of the luminosity of the burst in our targets. Assuming $L = 4 \pi r^2 \sigma_{\rm SB} T^4$, and half of the bolometric IR luminosity from  \citep[][]{SANDERS03} to be in the burst, we find that $T_{\rm dust}$ must be $< 160$~K. We consider a reasonable plausible range $T_{\rm dust} \sim 60{-}160$~K; given that the ammonia temperatures for the ``hot'' components lie in the intermediate part of this range, $T_{\rm dust} \sim 130$~K with $50\%$ uncertainty seems like a reasonable assessment.

As noted, $\kappa$ appears uncertain by a factor of $\sim 2$. Allowing a $\sim 30\%$ uncertainty in the dust-to-gas ratio, the overall uncertainty on the gas mass is likely $\sim 0.4{-}0.5$~dex.

For comparison, our assumptions yield an $\approx 30$ times lower gas mass than what one would calculate from the $350$~GHz light-to-gas-mass conversion of \citet{SCOVILLE16}. That is, we take the dust in these proto-SSCs to be more emissive and much hotter than typical dust in galaxies. Bearing this in mind, we consider that our gas masses are most likely to be underestimates. 

\subsubsection{Dynamical Masses} 

We estimate dynamical mass of each source via 

\begin{equation}
M_{\rm VT} = 892~\ell_{\rm FWHM}~\sigma_v^2~.
\end{equation}

\noindent Here $\sigma_v$ the measured velocity dispersion (in $\rm km/s$), $\ell_{\rm FWHM}$ is the full width half max deconvolved size of the source (both from Table \ref{tab:peaks}), and $M_{\rm VT}$ is the virial theorem-based dynamical mass in units of solar masses. The prefactor here assumes a density profile $\rho \propto r^{-2}$.

Based on this calculation, we find median dynamical mass $\log_{10} M_{\rm VT} [\rm{M}_\odot ] \sim 5.7$ and values in the range $\log_{10} M_{\rm VT} [\rm{M}_\odot ] \sim 5.1{-}6.4$.

This calculation assumes that the line widths are due to self gravity. It corresponds to an upper limit if the velocity dispersion includes some contribution from inflow, outflow, or material unassociated with the source. If our sources break into multiple components at higher resolution, as Sgr B2 does in the Milky Way, then we also expect $M_{\rm VT}$ to represent an overestimate. We estimate the systematic uncertainty due to unresolved substructure to be $\lesssim 0.3$~dex, with the sense that our virial masses will be overestimated because we somewhat overestimate the size and line width of the source.

\subsubsection{Comparison of Mass Estimates} 

Figure \ref{fig:masses} compares our mass estimates. Our sources show gas and stellar masses $\sim 10^4$ up to $\sim 10^6$~M$_\odot$. Just as the sizes that we measure are typical of young cluster sizes \citep[e.g.,][]{RYON17}, these masses resemble those seen for massive clusters in nearby starbursts \citep[e.g.,][]{WHITMORE03,MCCRADY05}. Our sources thus do meet the definition of young massive clusters suggested by, e.g., \citet{PORTEGIESZWART10} and \citet{LONGMORE14}.

Our observations suggest a range of gas richness for the targets, but the left panel of Figure \ref{fig:masses} shows that gas often contributes a large fraction of the mass. In all but four sources gas contributes $\gtrsim 50\%$ of the mass (and bear in mind that we are suspicious of the M$_\star$ for sources \# 11 and 12). The median gas mass fraction ($M_{\rm gas}/(M_{\rm gas}+M_\star)$ across the sample is $\sim 50\%$, though with significant uncertainties. 

The right panel of Figure \ref{fig:masses} shows that our virial mass estimates tend to exceed our combined star plus gas estimates by a factor of $\sim 2.5$. Given the uncertainties in the gas and stellar mass estimation, this still represents reasonable agreement. The dynamical mass estimate reinforces that these structures mostly contain $\sim 10^5{-}10^6$~M$_\odot$ in a region a few pc across, with large contributions from both stars and gas. The discrepancy between the two total mass estimates could, in principle, reflect out-of-equilibrium motions (outflows or inflows). Or it might indicate that the sources are in a non-virialized dynamical state, for example, the line widths might reflect blended motion of unassociated sources. Just as likely, the discrepancy between the virial and gas plus stellar masses between the large uncertainties in our mass estimates, especially ionizing photons absorbed by dust and our uncertainties in $\kappa$ and $T_{\rm dust}$.

\subsubsection{Density and Free Fall Time} 

Table \ref{tab:masses} also reports the total (gas plus stellar) surface density, volume density, and implied gravitational free fall time. These are calculated within the FWHM, so that in two dimensions we divide the mass by $2$ and divide by the area at FWHM. In three dimensions, we divide the mass by $3.4$ before dividing by the volume at the FWHM.

The median surface density is $\log_{10} \Sigma_{\rm gas+\star} [{\rm M_\odot~pc^{-2}}] \sim 4.5$ and values lie range $3.7{-}5.1$.  Recasting these values in terms of mass surface density from the edge to the center of the structure (i.e., converting units and dividing by 2), our calculations imply a median $\sim 3.4$~g~cm$^{-2}$ from the center to the cluster edge and a range $0.5{-}14.0$~g~cm$^{-2}$. These values resemble those found in the Milky Way for other regions of high mass star formation \citep[e.g., see][]{MCKEE07}. The high end of our range of measured surface densities approaches the $\sim 20$~g~cm$^{-2}$ maximum surface density (now through the whole object, not center to edge) for stellar systems found by \citet{HOPKINS10}. On average, these proto-clusters are a factor of $\sim 3$ of below this maximum surface density.

The median gas plus stellar volume density in our targets is $\log_{10} \rho_{\rm gas+\star}  [{\rm M_\odot~pc^{-3}}] \sim 4.0$ (range $3.4{-}4.5$) in units of $M_\odot$~pc$^{-3}$. This would correspond to a median $n_H\sim 10^5 {\rm cm}^{-3}$ if all of the material were in molecular gas. The gravitational free fall times implied by these densities will be $\log_{10} \tau_{\rm ff} \sim 4.9$ (range $4.5{-}5.3$) years.

Considering only the gas mass, the implied surface densities for our sources would correspond to a median $\sim 500$~mag (range 20{-}4200~mag) of $V$-band extinction for a Milky Way dust-to-gas ratio \citep{BOHLIN78}. This helps explain why most of our targets do not appear as distinct sources in the HST imaging.

The final column of Table \ref{tab:masses} quotes the escape velocity $\sqrt{2GM/r}$ calculated within the 3-d FWHM of the source using the gas plus stellar masses. Using the data reported in the tables:

\begin{equation}
\label{eq:vesc}
v_{\rm esc} = \sqrt{\frac{2 G \left( \frac{M_\star + M_{\rm gas}}{3.4}\right)}{\left( \frac{\ell_{\rm FWHM}}{2} \right)}}
\end{equation}

\noindent where factors of $3.4$ account for the fraction of mass inside the FWHM of a $3$-d Gaussian and a factor of $2$ converts from FWHM to HWHM. Again $\ell_{\rm FWHM}$ refers to the FWHM, deconvolved size of the source from Table \ref{tab:peaks}.

These $v_{\rm esc}$ for all of our sources exceeds the $\sim 10$~km~s$^{-1}$ sound speed expected for photoionized gas. As a result, the clusters should match the definition for young massive protoclusters from \citet{BRESSERT12}.

\subsection{Notes on Individual Sources}

As already mentioned, sources \#11 and \#12 sit in a complex region. Much of the $36$~GHz flux that we measure may not directly associated with these clusters. We will treat these two sources with caution when drawing conclusions about the population as a whole. As also mentioned, source \#6 has a high $36$~GHz to $350$~GHz ratio. This could imply a more evolved cluster. But source \#6 is also our faintest source and the formal uncertainty in $f_{\rm ff}$ is quite high (see Table \ref{tab:peaks}). More, the contrast with local background or extended features associated with the bright, nearby source \#5 is poorer than for our other targets.

Several sources show complex line profiles. Sources \#3 and \#4 represent the clearest examples, but sources \#2 and \#13 also show significant non-Gaussian structure. We might expect both inflow and outflow during the evolution of a protocluster, leading to P-Cygni or reverse P-Cygni profiles. Peak \#4 shows some indication of this. We might also expect some of our sources to break up into collections of smaller objects, as would be the case for Sgr B2. This might help explain the profiles in peaks \#2 and \#3. Some of the central dips seen in the profiles, e.g., \#3 and \# 13 could alternatively reflect absorption from colder, denser gas.

Based on a first look at even higher resolution ALMA imaging obtained during review of the paper, our sources appear to represent the main bright point sources even at $< 1$~pc resolution. Many sources are associated with smaller satellite sources, and the region around sources \#10, \#11, and \#12 contains at least $6$ discrete point sources, with our three candidate protoclusters the brightest. These data will be presented, when science ready, in a future paper.

\section{Discussion}
\label{sec:implications}

We identify $14$ candidate young super star clusters in the inner region of NGC~253. How much of the star formation in the burst can these sources account for? What can we say about the efficiencies and timescales associated with these sources? And given such intense concentrations of gas and young stars, what can we infer about feedback in these sources?

\subsection{Timescales and Relation to the Starburst}
\label{sec:timescales}

The central burst in NGC 253 forms $\sim 2$~M$_\odot$~yr$^{-1}$ \citep{LEROY15A,BENDO15}. How much of that can be attributed to these sources?

{\bf Fraction of Ionizing Photons and IR Luminosity From These Sources:} We estimate a total ionizing photon production of $Q_0 \sim 1.2 \times 10^{53}$~s$^{-1}$ from our targets, with half of this coming from sources \# 11 and \# 12. \citet{BENDO15} find $Q_0=3.2 \pm 0.2 \times 10^{53}$~s$^{-1}$ for the whole burst. Thus, our sources may produce between 20 and 40\% of the total ionizing photons in the burst. Accounting for absorption of ionizing photons by dust, which must be more significant in our targets than any less embedded population, would increase this fraction.

An analogous case holds for the bolometric luminosity. If 50\% of the total infrared luminosity measured by \citet{SANDERS03} arises from the nuclear region, then $L_{\rm TIR} \sim 1.8 \times 10^{10}$~$L_\odot$ for this region. We estimate the contribution of our sources to this value by taking the light-to-mass ratio of a ZAMS population to be $\Psi\equiv L_\star/M_\star \sim 1000$~L$_\odot$~M$_\odot^{-1}$ and adopting the $M_\star$ calculated from the $36$~GHz emission. In this case our clusters together contribute $\sim 17\%$ of the bolometric luminosity of the burst; neglecting sources \# 11 and \# 12 this drops to $\sim 9\%$. 

\textbf{Relevant Timescales:} Several distinct timescales should combine to produce our observations. First, the timescale for cluster formation is the time spent in this compact, gas-rich phase. \citet{SKINNER2015} find a typical timescale of $\sim 5 \tau_{\rm ff}$ for cluster formation. The forming clusters simulated by \citet{SKINNER2015} do evolve over this time. The phase in which gas is actively collapsing to make stars is even shorter, $1-2 \tau_{\rm ff}$, and the surrounding gas dispersed by $\sim 8 \tau_{\rm ff}$. 
They measure their $\tau_{\rm ff}$ averaged over their $r=10$~pc cloud, with $\tau_{\rm ff} \sim 0.5$~Myr. We consider smaller scales and find free-fall times $\sim 10^5$~yr (Table \ref{tab:masses}). Assuming the \citet{SKINNER2015} results to scale with the free fall time, this implies a visibility timescale for forming clusters of $\sim 8 \times 10^5$~yr.

Second, ionizing photon production declines rapidly after $\sim 3{-}5 \times 10^6$~yr. This should be the timescale to produce the overall $Q_0$ in the burst. Third, infrared or bolometric luminosity is produced over a longer timescale than ionizing photons, with a single stellar population still producing significant light for many tens of Myr.

This short cluster formation timescale, $\sim 10^6$~yr, implies that a large amount fraction of stars in the burst are born in clusters (see next section). Our observations do support the idea that the clusters are young. Below, we show that the total radial momentum in these clusters appears low relative to their stellar mass. This implies that feedback has not yet unbound the protoclusters, consistent with an age young enough that a large amount of supernovae have not yet gone off. The clusters also show signs of ionizing photon production from compact regions.  Still, these radial momentum limits and the presence of ionizing photons only place hard limits of $\lesssim 5{-}10$~Myr on the age of the clusters. The value of $\sim 8~\tau_{\rm ff} \sim 1$~Myr should be viewed as a key assumption motivated by theory \citep{SKINNER2015}.

\textbf{Are Most Stars in the Burst Born in Clusters?} The estimated timescale for cluster formation, $\sim 10^6$~yr, is $\sim 20{-}30\%$ of the timescale for ionizing photon production. If all stars are born in clusters, we expect $20{-}30\%$ of the ionizing photons coming from the burst at any given time to arise from still-forming clusters. In this case, our observations agree with a large fraction ($\sim 100\%$) of stars being born in the burst proceeding through a phase like what we see.

Equivalently, we can see that the clusters might supply order unity of the star formation simply from their masses. Neglecting sources \#11 and \#12, which have questionable ionization and/or emission mechanisms, we find a total mass within the clusters of $M_{\rm gas+\star} \approx 3 \times 10^6$~M$_\odot$, split approximately equally between gas and stars. Combining this total mass with the $8~\tau_{\rm ff} \sim 10^6$~Myr cluster formation timescale based on \citet{SKINNER2015}, then our observed sources can already account for more than the total $\sim 2$~M$_\odot$~yr$^{-1}$ SFR in the burst. This assumes continuous star formation at the time-average rate, leverages our assumed cluster formation timescale, adopts a 100\% gas to star conversion efficiency, and relies on our somewhat uncertain mass estimates. All of these assumptions likely break down in detail. But the calculation shows that the clusters represent a large fraction of the mass that the burst has likely formed over the last $\sim 10^6$~yr.

Observations and theory both predict a larger fraction of star forms born in clusters in regions of high star formation surface density \citep[e.g.,][]{KRUIJSSEN12,JOHNSON16,GINSBURG18B}. The nucleus of NGC~253 has one of highest star formation surface densities in the local universe. Finding $\sim 100\%$ of the stars to be born in clusters in this extreme environment thus qualitatively matches expectations.

\textbf{A Plausible Scenario:} Our observations appear consistent with a scenario in which most of the stars in the burst form in massive young clusters. The formation process last for $\sim 1$~Myr, after which feedback disperses the immediate natal cloud \citep[][]{SKINNER2015}. After this, the clusters remain present, but without an associated large gas reservoir. They will still be invisible in the near-IR, hiding behind the high overall extinction in the region. They will also continue to produce ionizing photons, contributing to the overall $Q_0$ in the burst inferred from free-free and radio recombination line emission \citep{BENDO15}. Meanwhile their corresponding {\sc Hii} regions would grow in size, fading in surface brightness and becoming much more difficult to pick up in our interferometric radio continuum maps. Eventually, they might evolve in analogs to the older, visible clusters seen at larger radii by \citet{FERNANDEZ09}.

We expect the strong feedback that drives the X-ray and molecular gas winds \citep[][]{STRICKLAND02,BOLATTO13A,WALTER17} to occur after the embedded young cluster phase that we observe. After $\sim 10$~Myr, many massive stars will explode as supernovae. These explosions may trigger both the hot and cold outflows. A scenario in which strong feedback occurs well after the embedded phase agrees with our observations, which show that the protocluster candidates are approximately gravitationally bound at 2~pc scales, with no evidence for high velocity line wings in their spectra. In this case the clusters that we observe now are not the immediate drivers of the outflows observed in \citet{BOLATTO13A} \citet{WALTER17}, and \citet{ZSCHAECHNER18}. They may, however, drive similar outflows in the future.

\textbf{Lower Mass Clusters:} We observe candidate protoclusters down to a combined gas plus stellar mass of $\log_{10} M_{\rm gas+\star} \sim 4.8$. The cluster mass function is often taken to have equal power per decade \citep[e.g.,][]{PORTEGIESZWART10}. Taking the cutoff for young massive clusters as $\sim 10^4$~M$_\odot$ \citep{PORTEGIESZWART10,LONGMORE14,BRESSERT12}, there may be as much mass in low mass, unidentified clusters as in the sources that we study. Likely many of these lower mass sources will be substructure still unresolved by our $1.9$~pc beam, analogous, e.g., to Sgr B2. In this case they would already be counted in our bookkeeping. But we also likely miss some peaks that remain blended at our resolution or show too weak a contrast against the background to be detected by our peak finding algorithm. As a result, there may be even more cluster formation in our region of interest than we observe here. This point should be addressed by ongoing higher resolution ALMA observations.

Of course, if twice as much mass --- and ionizing photons and bolometric luminosity --- are present in clusters outside our sources then our bookkeeping above breaks. This could indicate a longer timescale for cluster formation or it might reflect that we have systematically overestimated the mass of the clusters.

\textbf{Lower Limit:} As emphasized, the cluster formation timescale remains uncertain. Our mass estimates are also uncertain at the factor of two level. Given that all of the sources show 36~GHz flux and that at least $20\%$ of the ionizing photon production occurs in our sources, a reasonable limiting case is that the visibility lifetime for the clusters equals the $\sim 3{-}5$~Myr ionizing photon production time and that $\sim 20\%$ of the stars in the burst form in these structures. Even in this limiting case, the burst represents a prodigious cluster production factory, far more extreme than what we see around us in the Milky Way.

\subsection{Likely High Efficiency Per Free Fall Time}

We find $M_\star \sim M_{\rm gas}$ and free fall times $t_{ff} \sim 10^5$~yr based on the combined gas plus stellar mass in the clusters. Assuming that no mass has escaped from the cluster since its initial formation, then $M_\star \sim M_{\rm gas}$ implies an overall efficiency of $\sim 50\%$. That is, 50\% of the initial total mass is now in stars. Following the argument above that cluster formation occurs over $\sim 5~t_{ff}$, this implies an efficiency per free fall time of $> 10\%$, again assuming no mass loss. If we adopt a visible lifetime of $\sim 1$~Myr based on the fraction of ionizing photons seen in the sources, this would instead imply $\sim 5\%$ of the gas mass converted to stars per free fall time. 

Note that some mass-loss may already have occurred \citep[e.g., as might be expected following][]{THOMPSON16}, in which case the efficiency per free-fall time would be lower than we calculate. In order to have a $\sim 1\%$ efficiency per free fall time as is observed at larger scales \citep[e.g.,][]{KRUMHOLZ07,KRUMHOLZ12B,UTOMO18}, several times the currently observed gas must have already been expelled; however, this seems unlikely based on observed linewidths.

\subsection{High Infrared Opacity}

The deconvolved $350$~GHz peak brightness temperatures associated with our sources are high, often $\sim 10$~K and in a few cases $20{-}40$~K or more. We do not know the true dust temperature, but our arguments above suggest that it cannot be much more than $\sim 130$~K on average. In this case, the dust optical depth at $350$~GHz is already $\tau \sim 0.1$ in many of our compact sources.

{\bf Opaque at IR wavelengths:} For dust, $\tau \propto \nu^\beta$ with $\beta \sim 1.5{-}2$ in the far-infrared and submillimeter. Combined with the significant $\tau$ at $350$~GHz, this implies that dust continuum emission from our compact sources will be optically thick for wavelengths shorter than $\lambda \sim 200{-}300\,\mu$m. They will have a factor of $\sim 50{-}100$ higher optical depth at $\lambda \sim 100\,\mu$m compared to 350~GHz ($855\mu$m). This yields optical depths $\tau \sim 5{-}10$ at 100 $\mu$m, and much larger near the implied peak of the dust SED at $\sim 20{-}30\mu$m.

With such high optical depths, these cluster-forming structures might be expected to have large IR photospheres. The regions could appear much larger at IR wavelengths than at sub-mm wavelengths, so that resolving them is only possible with ALMA. Clumpy substructure might render this a more local effect, so that the gross morphology of the sources does not change, but the opacity effects must be present at some scale. This clumpy substructure might be expected from comparing the mean particle densities, $n\sim 10^5$~cm$^{-3}$, with the typical densities needed to excite the bright H$^{13}$CN~(4-3) and CS~(7-6) emission we measure, which requires densities $n\sim 10^7$~cm$^{-3}$.

{\bf Significant Infrared Radiation Pressure Force:} This high opacity at IR wavelengths also implies a strong radiation pressure force exerted by the cluster stars on the surrounding gas \citep{MURRAY10}. For spherical systems optically thick to stellar radiation, the stellar radiation creates an outward force $L_\star/c$, with $L_\star$ the bolometric luminosity. For systems that are  optically thick in the infrared, the reprocessed infrared light also contributes to this outward force. This force due to infrared radiation force exceeds the force associated with the primary stellar radiation by a factor equal to the Rosseland mean optical depth $\sim \tau_{\rm IR}$. The high $\tau_{\rm IR}$ in our clusters thus implies a strong radiation pressure force on the surrounding gas. 

{\bf Effect of Radiation Pressure on Cluster Formation:} How does this high radiation pressure affect cluster formation? Following \citet{MURRAY10} and \citet{SKINNER2015}, for a spherical system centered on a star cluster, the ratio of the IR radiation force to the gravitational force from the stars is 

\begin{eqnarray}
\label{eq:iredd}
f_{\rm Edd, IR} &=& \frac{\kappa_{\rm IR,gas} F_{\rm IR}/c }{G M_*/r^2} = \frac{\kappa_{\rm IR,gas } \Psi}{ 4 \pi c G},
\end{eqnarray}

\noindent where $F_{\rm IR}=L_\star/(4 \pi r^2)$ is the IR flux, assuming all starlight to be reprocessed into the IR. Here $\kappa_{\rm IR,gas}$ is the mass absorption coefficient {\em per unit gas mass}; note the difference from above where we discuss the mass absorption coefficient of dust alone. Thus $\kappa_{\rm IR,gas}$ includes both the dust properties and the dust-to-gas ratio. Here $\Psi\equiv L_\star/M_\star$ refers to the light-to-mass ratio of the central stellar population.

If $f_{\rm Edd, IR} > 1$ then the radiation pressure force exceeds gravity and we might expect the collapse to halt. \citet{SKINNER2015} demonstrated this using numerical radiation hydrodynamic simulations, showing that when $f_{\rm Edd, IR} > 1$ the SFE is limited to $\sim 50\%$.  Their simulations also showed that  the radiation field cannot limit collapse in turbulent, SSC-forming clouds when $f_{\rm Edd,IR}<1$ \citep[see also][]{Tsang17}.

The stellar population sets $\Psi$, while $\kappa_{\rm IR, gas}$ is set by dust properties and the dust abundance relative to gas. Adding gas to the system increases the total dust opacity, leading to a higher $\tau_{\rm IR}$ and more support from radiation pressure. But at the same time, adding gas to the system increases the weight of gas. Because these two effects balance, a high $\tau_{\rm IR}$ does not necessarily imply anything about force balance in the cluster \citep[though there can be an indirect dependence of $\kappa_{\rm IR, gas}$ on $\tau_{\rm IR}$ through the dust temperature][]{KrumholzThompson12}.

For a zero age main sequence with a Kroupa initial mass function, $\Psi \sim 2000 ~{\rm erg~s^{-1}~g^{-1}}$ $\sim 1000~L_\odot~M_\odot^{-1}$. For temperature range and dust abundance relevant to our clusters, \citet{SEMENOV2003} find a Rosseland mean opacity $\kappa_{\rm IR,gas} \lesssim 5~{\rm cm^2~g^{-1}}$. In this case $f_{\rm Edd} \lesssim 0.4$ and stellar gravity would exceed the IR radiation pressure force by a factor $\gtrsim2$. In this case, radiation pressure would help support the cloud against collapse, but not supply all of the support nor tear the cloud apart. Including gas self-gravity would only strengthen the effects of gravity relative to radiation pressure. 

This situation could change if $\Psi > 1000~L_\odot~M_\odot^{-1}$, e.g., due to a top-heavy IMF. Top-heavy IMFs have been claimed in 30 Doradus \citep[][]{SCHNEIDER18} and the proto-SSC in NGC~5253 \citep{TURNER17}. Alternatively, if the gas associated with the clusters has a higher than Galactic dust-to-gas ratio, or unusually opaque grains, $\kappa_{\rm IR}$ would be higher than assumed above.

We find virial masses within a factor of $\sim 2$ of $M_\star + M_{\rm gas}$. This supports a scenario in which the clusters are gravitationally bound in approximate equiibrium. It appears that radiation forces, though certainly enhanced by a high $\tau_{\rm IR}$, at most balance gravity at the present time, consistent with the expectations above. At present, we lack independent constraints on the dust to gas ratio, nor do we independently measure $L_\star$ and $M_\star$.

\subsection{Limits on Feedback From Observed Radial Momentum}

The correspondence between virial masses and $M_\star + M_{\rm gas}$ implies that gravity approximately balances the outward force in our clouds. Given enough time, both supernovae (SNe) and stellar winds can inject enough momentum to unbind the gas and drive a radial expansion. The contrast between the observed radial motions in our sources and the expected momentum injection from SNe and stellar winds provides additional indirect support for the idea that our sources are young.

{\bf Observed Limits on Radial Momentum:} Taking all motions to be radial and outward, the momentum per unit stellar mass for an expanding spherical system is

\begin{equation}\label{eq:radmom}
\frac{p_r}{M_\star}\equiv \sqrt{3}\sigma_v \frac{M_{\rm gas}}{M_\star}~.
\end{equation}   

\noindent This $p_r$ will be the maximum radial momentum compatible with an observed velocity dispersion $\sigma_v$ and gas mass $M_{\rm gas}$. Normalizing by $M_\star$ allows a direct comparison with input from SNe and stellar winds, which both scale with stellar mass.

We report $p_r/M_\star$ limits for our sources in Table \ref{tab:masses}. We find mostly $p_r/M_\star < 100 ~{\rm km~s^{-1}}$, with the largest value $\sim 400 ~{\rm km~s}^{-1}$ for source \#3, which has a high gas mass relative to its stellar mass and also a line profile suggestive of significant substructure (Figure \ref{fig:profiles}).

{\bf Momentum From Supernova Feedback:} Numerical simulations considering clustered SNe exploding in an inhomogenous medium find a momentum injection per SN (after cooling and shell formation) of $\sim 10^5 M_\odot ~ {\rm km~s^{-1}}$ \citep[e.g.][and references therein]{KIM2017}. For a Kroupa IMF, with roughly one supernova per 100~$M_\odot$ formed, we expect $p_\star/M_\star \sim 10^3 ~{\rm km~s^{-1}}$ at late times ($\sim 10$ Myr). This is an order of magnitude higher than what we observe for most sources, suggesting that supernovae have not yet had a significant effect on internal motions.

The large values of $p_\star/M_\star$ are associated with a long timescale, $t_{\rm SN} \gtrsim 10$~Myr. Spreading $p_\star/M_\star \sim 10^3 ~{\rm km~s^{-1}}$ across this $t_{\rm SN}$, the mean momentum injection rate, $(p_\star/M_\star)/t_{\rm SN}$, may not exceed the gravitational force, $GM_{\rm gas + \star}M_{\rm gas}/(M_\star r^2) \sim (p_r/M_\star)/t_{\rm ff}$, especially at early times. 

This argument does not preclude any supernovae having gone off. \citet{KORNEI09} note the presence of iron lines in their SSC (our source \#5). Our limits should be read as indicating that the observed momentum in the cluster does not reflect a set of clustered --- in space and time --- supernovae explosions with sufficient intensity to unbind the clusters. This comparison then places a relatively weak constraint on the object age to be $\lesssim 10$~Myr.

{\bf Momentum From Stellar Winds:} For stellar wind feedback, the pressure-driven bubble solution of \citet{WEAVER1977} yields a ratio of shell momentum to central cluster mass of 

\begin{equation}
\frac{p_\star}{M_\star}= 65 ~{\rm km~s^{-1}} \dot E_{w,34}^{4/5} n_5^{1/5} M_{\star,5}^{-1/5} t_5^{7/5}~.
\end{equation}

\noindent Here $\dot E_{w,34}$ is the average wind luminosity injected per $M_\odot$ of stars in units $10^{34} {\rm erg~s^{-1}}$; $n_5$ is the mean hydrogen density in units $10^5 ~{\rm cm}^{-3}$; $M_{\star,5}$ is the cluster mass in units $10^5 M_\odot$; and $t_5$ is the cluster age in units $10^5$yr. From Starburst99, $\dot E_{w,34}=1$ \citep{LEITHERER99}.

This calculation also predicts the radius of the wind-driven bubble. If there are no energy losses and gravity is negligible,

\begin{equation}
r_b = 3 ~{\rm pc} \dot E_{w,34}^{1/5} n_5^{-1/5} M_{\star,5}^{1/5} t_5^{3/5}~.
\end{equation}

\noindent Though we are not yet in a position to measure the relative structure of the ionized gas, molecular gas, and dust, we do not expect $r_b$ to exceed our observed source size. In that case, the stellar winds would have cleared out the cold gas.

If the sources in NGC 253 are young ($t_5 < 1$), both the momentum injection and bubble size would remain below the observed limits for most sources even without losses. But if the sources are closer to $\sim 1$ Myr in age, as seems likely given their inferred $t_{\rm ff}$, then the predicted bubble radius and wind momentum input may significantly exceed our observed limits.

Again, it seems likely that the momentum injection from the wind has been balanced by gravity. Comparing the predicted momentum input rate from a wind-blown bubble, $(p_\star/M_\star)/t \propto t^{2/5}$ to the gravitational force, $(p_r/M_\star)/t_{\rm ff}$, this calculation suggests that the force from winds should exceed that of gravity for $t\sim10^6$~yr. Again, the lack of strong signatures of gas expulsion argue that our sources are young, $\lesssim 10^6$~yr. 
In this case, the effective wind luminosity must be reduced below the expected input value, either by mixing and cooling or by other processes. This is reasonable based on the low X-ray emission observed in somewhat more evolved systems, where the energy in hot gas seems to be far below the value nominally expected from winds, consistent with a reduction in $\dot E_{w,34}$ well below unity \citep[e.g.][]{Harper-Clark09,LOPEZ2011,ROSEN2014}.

\section{Summary}

We present new, $\sim 2$~pc resolution ALMA observations of the 350~GHz dust continuum emission from the innermost region of NGC~253, the nearest nuclear starburst. This imaging reveals $14$ compact, bright sources. The association of these dust emission peaks with VLA $36$~GHz continuum suggests that they already host significant populations of massive young stars \citep[many also appear in the 23~GHz images of][]{ULVESTAD97}. Despite this, the heavy extinction towards the burst renders all but one of them indistinguishable in near-IR emission. The one known source has previously been identified as a young super star cluster by \citet{WATSON96} and \citet{KORNEI09}. Studying these sources, we find:

\begin{enumerate}
\item {\bf Massive, Compact Sources:} We measure source sizes from the ALMA 350~GHz continuum (Table \ref{tab:peaks}). We estimate gas masses based on dust emission, and calculate stellar masses assuming that the observed 36~GHz continuum is free-free emission from a zero-age main sequence population (Table \ref{tab:masses}). 

We find sizes of a few pc (FWHM) and estimate total masses $M_{\rm gas} + M_\star \gtrsim 10^5$~M$_\odot$. We also estimate dynamical masses from the measured sizes and line widths and assuming virialization. The virial masses are typically $\sim 2.5$ times higher than our $M_{\rm gas} + M_\star$ estimates, which represents reasonable within the substantial uncertainties on the mass estimation.

\item {\bf Likely Young Super Star Clusters:} These masses and sizes resemble those of young massive clusters seen in the Milky Way and other galaxies \citep{PORTEGIESZWART10,LONGMORE14}. More, these masses and radii imply escape speeds $>16$~km~s$^{-1}$. This is larger than the sound speed of photo-ionized gas, $\sim 10$~km~s$^{-1}$, so that the sources also match the criteria for young massive protoclusters laid out by \citet{BRESSERT12}. Clusters in this mass range are often referred to as super star clusters.

\item {\bf Still in the Process of Formation:} Our estimates of the gas and stellar mass, while uncertain, suggest that gas still contributes a large fraction of the total mass in these objects (Table \ref{tab:peaks}). We observe that the dust emission coincides with H$^{13}$CN (4-3) and CS~(7-6) emission, both tracers of dense, excited gas. Thus, many of these objects seem likely to still be in the process of formation. 

\item {\bf Short Free-Fall Times and High Efficiency:} The free fall times implied by the gas plus stellar masses of our sources is short, $\tau_{\rm ff} \sim 10^5$~yr. Given theoretical expectations of $\sim 8~\tau_{\rm ff}$ for the cluster formation timescale \citep{SKINNER2015}, this implies that the sources are young. Stars typically represent half of the mass in our sources, implying both a high net efficiency and a high efficiency per free fall time. Note that this  statement does not take into account possible mass loss. Any mass lost from the system would decrease the both the net star formation efficiency and the efficiency per free fall time.

\item {\bf A Large Fraction of Stars Form in these Sources:} At least $20\%$ of the ionizing photon production in the burst appears associated with these sources. This represents a firm lower limit on the fraction of stars that form in such sources. If the cluster formation timescale is short compared to the time for stars to produce ionizing photons, then an even larger fraction of star formation may proceed through this phase. Accounting for a short cluster formation timescale and the possibility of lower mass, still-unidentified clusters, order $\sim 100\%$ of the stars in the burst could be produced in these sources. This number remains uncertain due to uncertainties in both the mass estimates and timescales.

\item {\bf Opaque in the Infrared:} These sources have high brightness temperature. Given plausible dust temperatures, they also have moderate ($\tau \sim 0.1$) optical depths at sub-millimeter wavelengths. This implies large optical depths near the peak in the infrared near the likely peak of their spectral energy distributions. In turn, this implies a significant infrared radiation pressure force. Given the reasonable agreement between the dynamical masses and our other estimates of $M_{\rm gas}+M_\star$, this force may help support the clouds but is not unbinding them. This agrees with theoretical expectations.

\item {\bf Young Based on Large Gas Fraction and Being Approximately Bound:} Our sources retain a large fraction of their mass in gas (as evidenced by the dust continuum). They also appear to be approximately gravitationally bound. We also calculate limits on the radial momentum in our sources and compare them to expectations from supernova and stellar wind feedback. Our sources have lower radial momentum and smaller sizes than expected from either clustered supernova or stellar winds acting over many Myr, though losses in wind energy may be important. All of these pieces of evidence suggest that the sources are young enough that feedback has not managed to unbind the gas from the cluster.

\end{enumerate}

Given the brightness of these sources, ALMA and the Jansky VLA both offer the prospect for even more detailed detailed follow up. Higher resolution dust observations are already underway, as is the construction of full radio-to-mm SEDs for each source \citep[building on][and leveraging new ALMA and VLA work]{ULVESTAD97,MOHAN05}.

It will also be important to link these structures larger context of the burst. In the Milky Way's Central Molecular Zone, star formation has been linked to the orbital paths of individual clouds \citep[e.g.,][]{KRUIJSSEN14B}. The linear distribution of the sources we see suggests an underlying bar-like structure \cite[see][]{LEROY15A,PAGLIONE04} or loosely wound arms. It may be possible to link this structure to the triggering of star formation. More generally, we do not see clear analogs for these structures in the Milky Way. This might be because NGC~253 sits at a different part of some long term nuclear fueling cycle \cite[e.g.,][]{KRUMHOLZ17}. A more detailed comparison of the two systems \citep[building on][]{SAKAMOTO11} could help reveal the overall triggers and likely duty cycle of the burst. This might also help reveal the fate of the proto-clusters after they disappear from our ALMA and VLA imaging, and perhaps link them to clusters seen on larger scales outside the area we study \citep[][]{FERNANDEZ09}.

\acknowledgments

We thank the anonymous referee for a constructive report that improved the paper. We also thank Gerhardt Meurer and Mark Krumholz for useful feedback during revision. 

This paper makes use of the following ALMA data: ADS/JAO.ALMA\#2015.1.00274.S. ALMA is a partnership of ESO (representing its member states), NSF (USA) and NINS (Japan), together with NRC (Canada), NSC and ASIAA (Taiwan), and KASI (Republic of Korea), in cooperation with the Republic of Chile. The Joint ALMA Observatory is operated by ESO, AUI/NRAO and NAOJ. The National Radio Astronomy Observatory is a facility of the National Science Foundation operated under cooperative agreement by Associated Universities, Inc. 

The work of AKL is partially supported by the National Science Foundation under Grants No. 1615105, 1615109, and 1653300. The work of ADB is supported in part by the NSF under grant AST-1412419. The work of ECO is supported by the NSF under grant AST-1713949. The work of TAT is supported in part by NSF Award 1516967.

We acknowledge the usage of the Extragalactic Distance Database\footnote{\url{http://edd.ifa.hawaii.edu/index.html}} \citep{TULLY09}, the HyperLeda database\footnote{\url{http://leda.univ-lyon1.fr}} \citep{MAKAROV14}, the NASA/IPAC Extragalactic Database\footnote{\url{http://ned.ipac.caltech.edu}}, and the SAO/NASA Astrophysics Data System\footnote{\url{http://www.adsabs.harvard.edu}}.

\facilities{ALMA, VLA, HST}

\software{{\tt CASA} \citep{CASA07}, {\tt IDL}, CPROPS\footnote{\url{https://github.com/akleroy/cpropstoo}} \citep{ROSOLOWSKY06,LEROY15A}}

\bibliography{akl}

\end{document}